\documentclass[10pt,preprint]{aastex62}

\graphicspath{{plots/}}

\usepackage{mathrsfs, amsmath}
\usepackage{dblfloatfix}
\usepackage{csquotes}
\usepackage{dblfloatfix}
\usepackage{fixltx2e}
\usepackage{url}
\usepackage{natbib}

\usepackage{graphicx}
\usepackage{enumitem}

\newcommand{\Iminus}{|I_-|}

\newcommand{\Iminuslpf}{\text{I}_\text{m, lf}}
\newcommand{\imlpf}{\text{im}_\text{lf}}
\newcommand{\lpf}{F_\text{LP}}
\newcommand{\appropto}{\mathrel{\vcenter{\offinterlineskip\halign{\hfil$##$\cr\propto\cr\noalign{\kern2pt}\sim\cr\noalign{\kern-2pt}}}}}

\accepted{July 6, 2018}
\submitjournal{AJ}

\shorttitle{Fast Coherent Differential Imaging}
\shortauthors{Gerard, B. L. et al.}

\begin{document}
%
\title{Fast Coherent Differential Imaging on Ground-Based Telescopes using the Self-Coherent Camera}
%

\correspondingauthor{Benjamin L. Gerard}
\email{bgerard@uvic.ca}

\author[0000-0003-3978-9195]{Benjamin L. Gerard}
\affil{University of Victoria, Department of Physics and Astronomy \\
3800 Finnerty Rd \\
Victoria, V8P 5C2, Canada}
\affiliation{National Research Council of Canada, Astronomy \& Astrophysics Program \\
5071 West Saanich Rd\\
Victoria, V9E 2E7, Canada}

\author[0000-0002-4164-4182]{Christian Marois}
\affiliation{National Research Council of Canada, Astronomy \& Astrophysics Program \\
5071 West Saanich Rd\\
Victoria, V9E 2E7, Canada}
\affiliation{University of Victoria, Department of Physics and Astronomy \\
3800 Finnerty Rd \\
Victoria, V8P 5C2, Canada}

\author{Rapha\"el Galicher}
\affiliation{Lesia, Observatoire de Paris, PSL Research University, CNRS, Sorbonne Universit\'es, Univ. Paris Diderot \\
UPMC Univ. Paris 06, Sorbonne Paris Cit\'e, 5 place Jules Janssen \\
92190 Meudon, France}
\begin{abstract}
Direct imaging and spectral characterization of exoplanets using extreme adaptive optics (ExAO) is a key science goal of future extremely large telescopes and space observatories. However, quasi-static wavefront errors will limit the sensitivity of this endeavor. Additional limitations for ground-based telescopes arise from residual AO-corrected atmospheric wavefront errors, generating millisecond-lifetime speckles that average into a halo over a long exposure. A solution to both of these problems is to use the science camera of an ExAO system as a wavefront sensor to perform a fast measurement and correction method to minimize these aberrations as soon as they are detected. We develop the framework for one such method based on the self-coherent camera (SCC) to be applied to ground-based telescopes, called Fast Atmospheric SCC Technique (FAST). We show that with the use of a specially designed coronagraph and coherent differential imaging algorithm, recording images every few milliseconds allows for a subtraction of atmospheric and static speckles while maintaining a close to unity algorithmic exoplanet throughput. Detailed simulations reach a contrast close to the photon noise limit after 30 seconds for a 1\% bandpass in H band on both 0$^\text{th}$ and 5$^\text{th}$ magnitude stars. For the 5th magnitude case, this is about 110 times better in raw contrast than what is currently achieved from ExAO instruments if we extrapolate for an hour of observing time, illustrating that sensitivity improvement from this method could play an essential role in the future detection and characterization of lower mass exoplanets.
\end{abstract}
\section{Introduction}
\label{sec: intro}
Direct imaging of exoplanets on ground-based telescopes is currently limited in sensitivity by both dynamic and quasi-static wavefront errors. Dynamic wavefront errors arise from the residual aberrations left uncorrected by an adaptive optics (AO) system, seen by the science camera as short-lived atmospheric speckles to longer-lived dome seeing effects, ultimately averaging to into a halo over a long exposure. Quasi-static wavefront errors instead arise from uncalibrated imperfections in the polishing and reflectivity of the optics in an instrument as well as from slower dome seeing effects. These aberrations produce a speckle pattern on the science camera. The pattern is quasi-static in nature due to slower thermal and flexure-related changes in the instrument and/or variable dome conditions, ultimately requiring daily or often hourly calibrations.

In the current paradigm of ground-based exoplanet imaging, residual AO atmospheric speckles and quasi-static speckles are respectively optically addressed as follows:
\begin{itemize}
\item by using the highest performing, ``extreme'' AO (ExAO) systems that produce a residual wavefront error of typically less than about 100 nm rms in H band,
\item mainly by using high quality, few nm rms optics in the science path \citep[e.g.,][]{gpi_fresnel}, although ongoing efforts are in place to calibrate low-order aberrations with the deformable mirror (DM) of an ExAO system \citep[e.g.,][Lamb, M., Norton, A. P., private communication]{zwfs}.
\end{itemize}
Subsequently, two main post-processing strategies are then used to further attenuate the speckle pattern:
\begin{enumerate}
\item spectral differential imaging \citep[SDI,][]{sdi1,sdi2,sdi3}, where images are recorded at different wavelengths in either multiple narrow bands or with an integral field spectrograph (IFS) to distinguish magnification of speckles as a function of wavelength vs. an exoplanet's fixed position, and
\item angular differential imaging \citep[ADI,][]{marois_phd, adi}, where the telescope's pupil orientation is fixed while tracking a star such that the relative position angle of an exoplanet rotates while the position angle of the telescope optics remains fixed.
\end{enumerate}
Although these two observing techniques are now often combined with advanced post-processing methods, such as a least-squares, singular value decomposition, and/or principal component analysis approaches \citep[e.g.,][]{loci, sosie, klip, devaney}, SDI and ADI are still fundamentally limited by chromaticity \citep[e.g.,][]{gpi_fresnel} and temporal stability \citep[e.g.,][]{marois_phd, sphere_stability}, respectively.

Atmospheric speckles provide a further limitation to temporal stability. \cite{speckle_lifetime} studied the statistics of atmospheric speckles in an AO-corrected coronagraphic image. In the absence of quasi-static aberration, if images are photon noise-limited so that atmospheric speckle amplitudes are consistently below the photon noise limit in individual exposures, the contrast (see appendix \ref{sec: setup} for a formal definition) will scale proportional to $t^{-0.5}$, where $t$ is the exposure time. If atmospheric speckle amplitudes are instead above the photon noise limit in individual exposures, contrast will remain flat when $t<\tau_\text{spec}$, where $\tau_\text{spec}$ is the speckle lifetime. If the image is dominated by atmospheric speckles and $t>\tau_\text{spec}$, contrast will scale proportional to $(t/\tau_\text{spec})^{-0.5}$ as atmospheric speckles average into a halo that is always above the photon noise limit. \cite{speckle_lifetime} show that the speckle lifetime scales with the atmospheric clearing time of a telescope aperture, which for 10-m class telescopes is typically on the order of a fraction of a second. This timescale will thus motivate our adoption of a fast (i.e., integrating on timescales less than the speckle lifetime) measurement and subtraction strategy.

A family of solutions that can be used to correct for both quasi-static and residual atmospheric aberration, separate from ADI and/or SDI, is called coherent differential imaging (CDI). CDI uses the science camera of an ExAO system to measure and subtract coherent speckles from the star without removing any incoherent exoplanet light. Over the past decade, a number of CDI techniques have been proposed, developed, and tested in the lab \citep[e.g.,][]{guyon_cdi, speckle_nulling, scc_orig, efc, wallace, phase_diversty}. These algorithms are generally designed for stable space telescope systems and require many iterations to converge at a minimum contrast. However, some CDI algorithms have recently been demonstrated on-sky, including:
\begin{itemize}
\item \cite{scexao_nulling} present the first on-sky CDI demonstration, using the speckle nulling algorithm originally proposed by \cite{speckle_nulling}. This process involved stepping through the unknown phase of each speckle in order to minimize the focal plane intensity in a region of interest, ultimately converging after 20 iterations of five second exposure frames. This iterative process limited contrast improvement to the first bright diffraction ring at 2.3 $\lambda/D$ (where $\lambda$ is the wavelength of light and $D$ is the telescope diameter), a relatively stable quasi-static feature for these timescales on the order of a minute. Residual atmospheric speckles and quasi-static speckles at higher spatial frequencies were not correctable at these frame rates, ultimately yielding a factor of 3 contrast improvement.
\item \cite{cdi_bottom} demonstrate a CDI approach using phase-shifting interferometry, where one path in a Mach-Zehnder interferometer is advanced through four piston offsets between 0 and 3$\pi$/2 radians. This approach is demonstrated on-sky at Palomar to present the first CDI-based detection of a substellar companion, the brown dwarf HD 49197b. Again, because this algorithm is iterative in nature, measurements and corrections were only capable on timescales longer than a few seconds. Correction of residual atmospheric and quasi-static speckles that generate temporal instability on shorter timescales was not feasible, ultimately yielding a factor of about 1-4 contrast improvement depending on different post processing methods that were used.
\end{itemize}
In this paper, we develop the framework for another CDI approach called the self-coherent camera \citep[SCC,][]{scc_orig,scc_lyot} to be applied to ground-based telescopes in order to correct both atmospheric and quasi-static aberration, hereafter referred to as ``FAST'' (for ``Fast Atmospheric SCC Technique''). In principle, our FAST approach is no longer limited by temporal stability as long as images are acquired faster than the dominant atmospheric and/or quasi-static speckle lifetime. In \S\ref{sec: fast} we present the foundation for our new FAST solution, including an optimized focal plane mask (FPM), called the SCC FPM (\S\ref{sec: scc_fpm}), long exposure simulations using the SCC FPM that motivate the potential gain of a FAST strategy (\S\ref{sec: long_exp}), and a post-processing strategy to measure and subtract coherent stellar speckles (\S\ref{sec: est_pin_psf}). In \S\ref{sec: sim} we present numerical detailed simulations of FAST performance, and then provide concluding remarks in \S\ref{sec: conclusion}. The appendices include a review the main principles of the SCC (\ref{sec: scc}, including post-processing algorithms in \ref{sec: post_processing} and photon noise propagation in \ref{sec: phlim}) and the parameters and assumptions used in our simulations (\ref{sec: setup}). At this time we do not yet consider chromaticity/broadband solutions, instrument-specific simulations, DM control, or a laboratory demonstration.
\section{The FAST Solution}
\label{sec: fast}
In this section we will both present and motivate the need for a FAST solution, which includes our new proposed SCC FPM and a post-processing subtraction strategy designed for exposures on the order of milliseconds. Simulations using the SCC FPM for a 30 second exposure will also illustrate the limitations of typical ground-based CDI observations due to residual atmospheric turbulence.
\subsection{The SCC FPM}
\label{sec: scc_fpm}
The conventional SCC design (see appendix \ref{sec: scc}) will only work on long exposures to measure quasi-static aberration. This restriction is a result of the amount of light transmitted through the pinhole in the Lyot plane. All previous SCC papers have focused on correction of quasi-static aberration on the order of a few nanometers rms because of this limitation; here, we now address wavefront control with the SCC more generically, by measuring and correcting for speckles on all time scales, including millisecond-lifetime atmospheric speckles. 

Assuming that we want to detect and subtract speckles as soon as they are visible and before they disappear, when a few photons are recorded for a speckle we ideally want the modulation amplitude to be at least as large as the speckle amplitude that we want to remove, requiring that the envelope of the pinhole point spread function (PSF) match the amplitude and power law of the stellar speckles. If this requirement is not met, fringes on a speckle of interest will be detected at a relatively lower signal-to-noise ratio (SNR) than the speckle, which will generate a noisier/biased calibration. Integrating longer will increase the fringe SNR to mitigate this problem, but this integration time must always in principle be shorter than the speckle lifetime; otherwise a sufficient SNR measurement of a speckle's fringes can never be made before the speckle disappears. 

\begin{figure}[!h]
\centering
\includegraphics[width=1.0\textwidth]{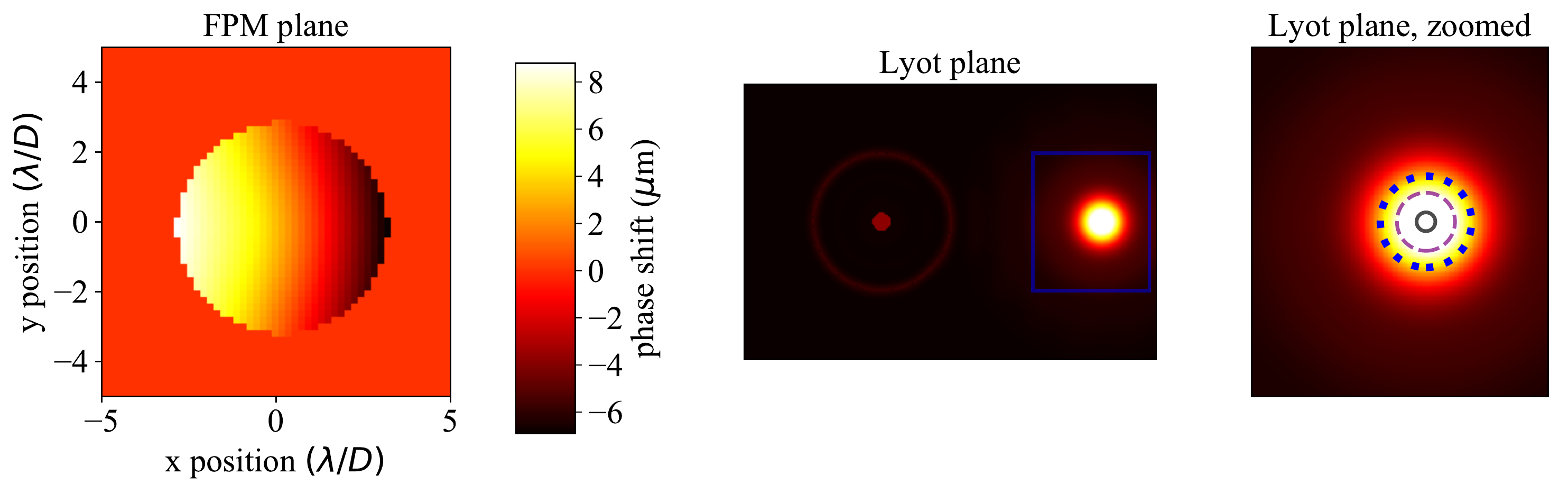}
\caption{Left: instead of using an amplitude FPM, a phase shift is applied in the focal plane, called the SCC FPM. Middle: the Lyot plane intensity after using the SCC FPM. The main pupil is on the left, and the off-axis pupil created from the SCC FPM is on the right. Right: a zoomed region of the blue rectangle in the middle panel. The dashed purple circle shows the theoretical minimum size of this off-axis pupil copy coming from the 6 $\lambda/D$ diameter SCC FPM compared to our measured FWHM (the dotted blue circle), illustrating that we are within a factor of about 2 from this limit. The solid black circle shows the size of the SCC pinhole for comparison, still well below the diffraction limit (the FPM inner working angle for a diffraction limit corresponding to the maximum pinhole size is is 9.3 $\lambda/D$). The integrated intensity in the Lyot pinhole with this FPM is increased by a factor of about $3.2\times10^5$ compared to the normal (unapodized) Lyot coronagraph.}
\label{fig: fpm_lyot}
\end{figure}

We ultimately found that the most successful way to get enough light through the pinhole was to design a new coronagraph, called the SCC FPM, which is outlined below. We replace the amplitude mask of the FPM in a simple Lyot coronagraph and instead apply a tilt to separations between zero and three $\lambda/D$, creating an off-axis pupil copy in the Lyot plane. This FPM acts as a spatial filter for the downstream off-axis pupil copy, filtering only low order modes ($\le3$ cycles/pupil). After applying a grid search of amplitude and width parameters for various additional low order aberrations, we found the best modulation amplitude results when using a 2.9 $\mu$m amplitude, 4.7 $\lambda/D$ full width at half maximum (FWHM) symmetric, two-dimensional Gaussian, shown in combination with the FPM tilt in Figure \ref{fig: fpm_lyot} along with the corresponding Lyot plane intensity. 
The right panel of Figure \ref{fig: fpm_lyot} shows the diffraction limit of a 6$\lambda/D$ diameter FPM (i.e., for a telescope with diameter $D/6$), and illustrates that our best solution is within a factor of 2 from this limit. Although our SCC FPM solution redistributes more light towards the center of the off-axis pupil, we are not actually ``shrinking'' this pupil; this would require either a smaller telescope and/or a mask that completely removes the smallest spatial scales of the complex electric field in the focal plane. 

\begin{figure}[!hb]
\centering
	\begin{minipage}[b]{0.35\textwidth}
		\begin{center}
		\includegraphics[width=1.0\textwidth]{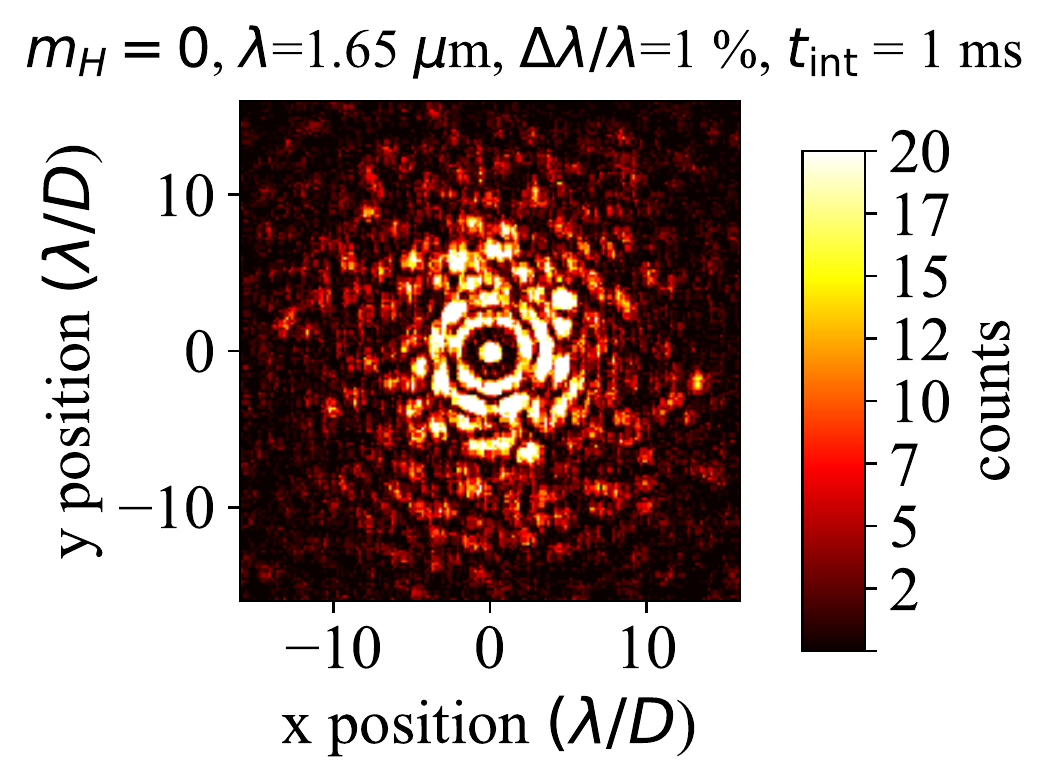}
		(a)
		\end{center}
	\end{minipage}
	\begin{minipage}[b]{0.6\textwidth}
		\begin{center}
		\includegraphics[width=1.0\textwidth]{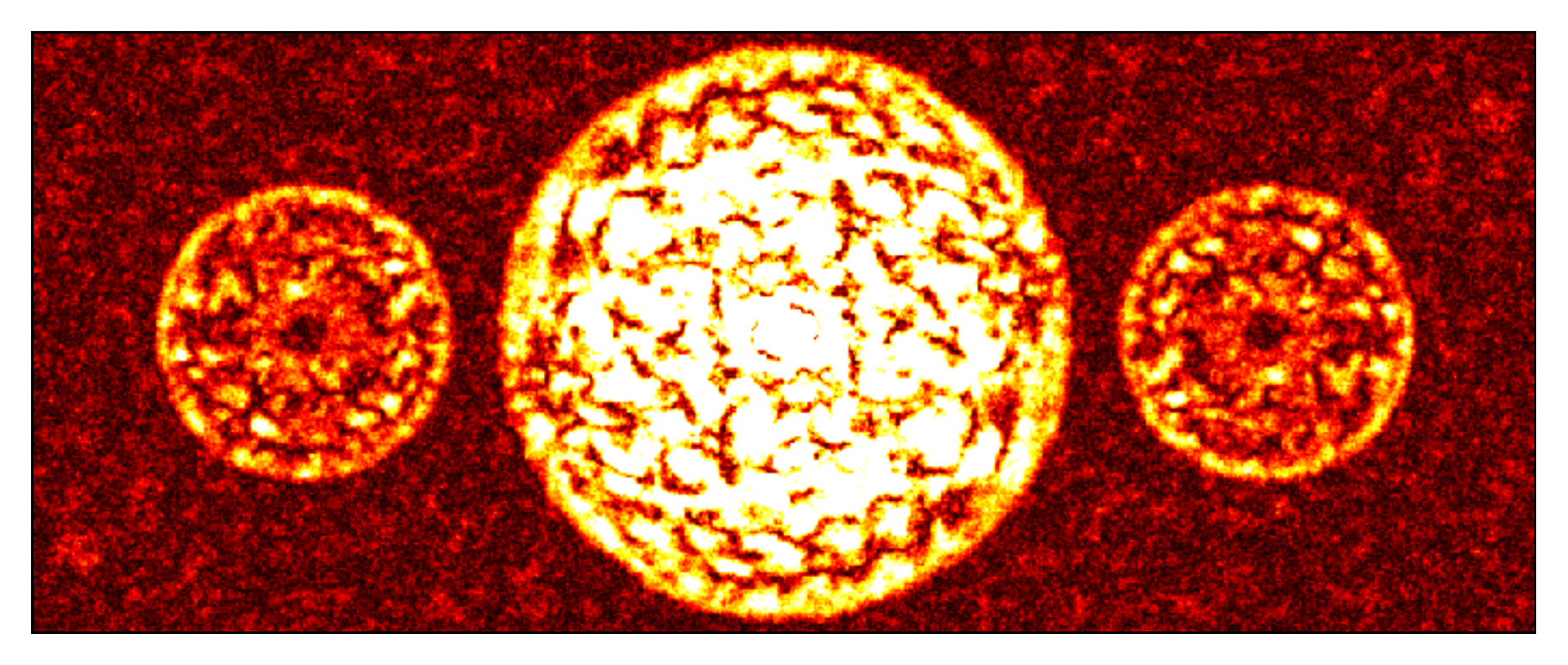}
		(b)
		\label{fig: b}
		\end{center}
	\end{minipage}	
\caption{(a) the SCC image, with photon noise, for a 1 ms exposure of a $m_H=0$ star using the SCC FPM design shown in Figure \ref{fig: fpm_lyot} and the wavefront parameters described in appendix \ref{sec: setup}. (b) The MTF of (a), illustrating that fringes (corresponding to the higher spatial frequency side-lobes in this image), are detected above the background photon noise floor.}
\label{fig: ms_exposure}
\end{figure}
A significant amount of light is still blocked from going through the pinhole. The pinhole cannot be made larger due to the limitations of equation \ref{eq: pinhole_size} (which requires that the 1.22 $\lambda/d$ minimum of the pinhole PSF lie outside the AO control region, where $d$ is the pinhole diameter remapped to the entrance pupil). This maximum pinhole size is also impossible to reach with our 6$\lambda/D$ diameter SCC FPM design, illustrated in the right panel of Figure \ref{fig: fpm_lyot}; the diffraction limit is $D/6$ whereas the largest pinhole size is (from equation \ref{eq: pinhole_size}) $D/18.5$ (i.e., reaching this diffraction limit would require an 18.5 $\lambda/D$ diameter FPM, or a 9.3 $\lambda/D$ inner working angle). However, despite these limitations, using the SCC FPM design in Figure \ref{fig: fpm_lyot} a sufficient amount of light is transmitted through the pinhole to detect fringes for exposures on the order of milliseconds; the image and modulation transfer function (MTF) for a 1 ms exposure of a $m_H=0$ star using the new SCC FPM design are shown in Figure \ref{fig: ms_exposure}, illustrating that the higher spatial frequency MTF sidelobes are detected above the photon noise floor. 
\subsection{Long Exposure Simulations and Analysis}
\label{sec: long_exp}
\begin{figure}[!hb]
\centering
	\begin{minipage}[b]{0.32\textwidth}
		\begin{center}
		\includegraphics[width=1.0\textwidth]{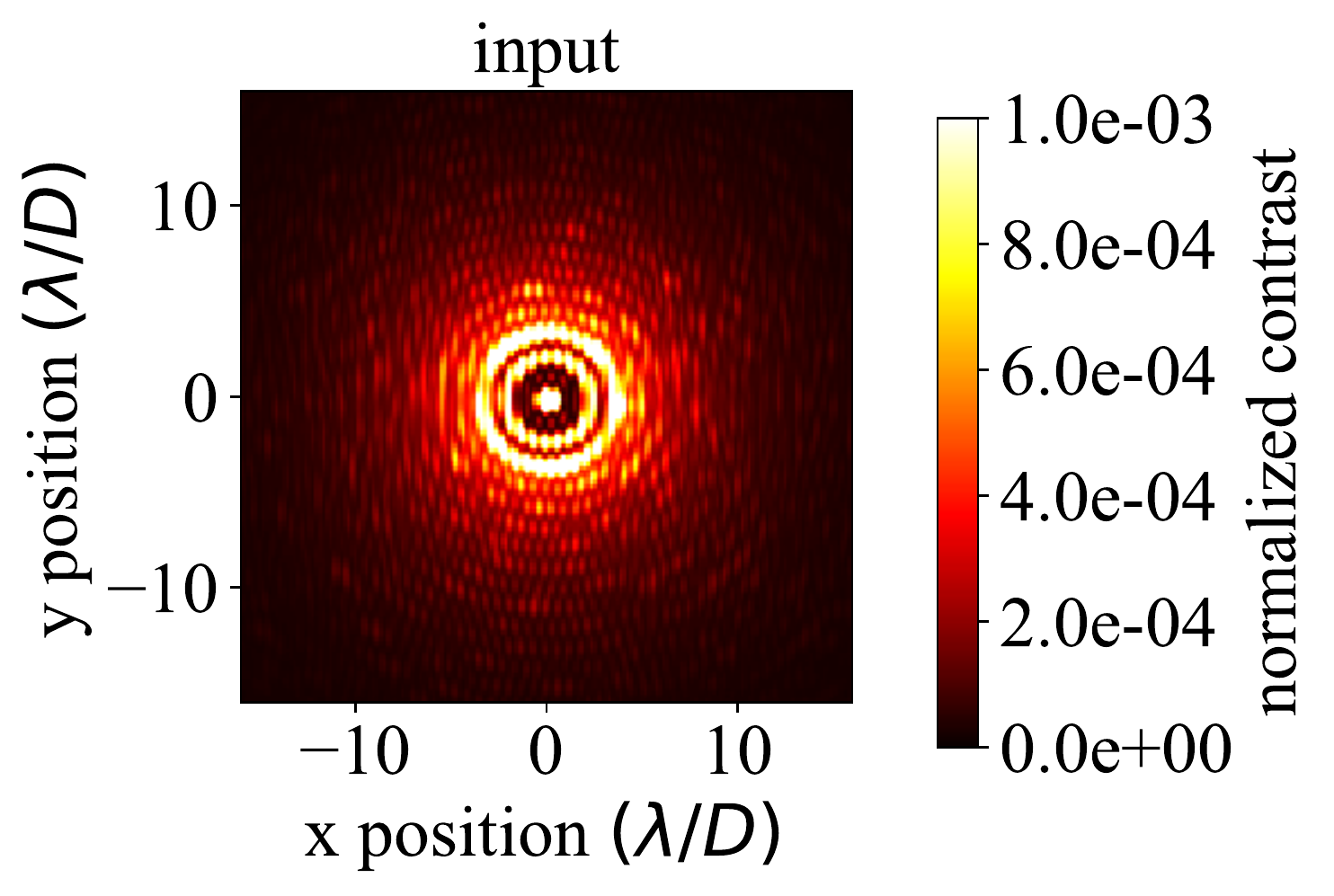}
		(a)
		\label{fig: a}
		\end{center}
	\end{minipage}
	\begin{minipage}[b]{0.32\textwidth}
		\begin{center}
		\includegraphics[width=1.0\textwidth]{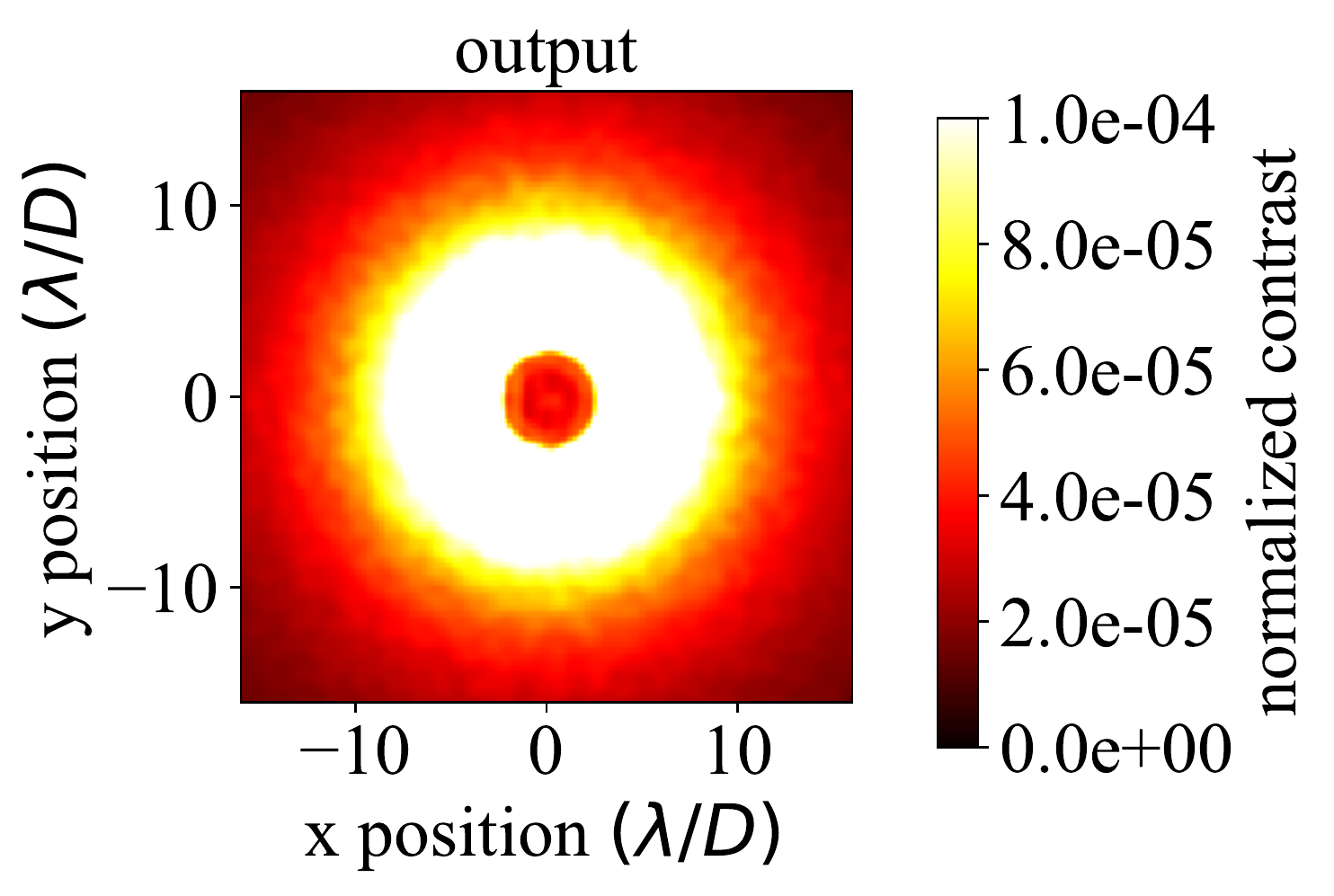}
		(b)
		\label{fig: b}
		\end{center}
	\end{minipage}
	\begin{minipage}[b]{0.32\textwidth}
		\begin{center}
		\includegraphics[width=1.0\textwidth]{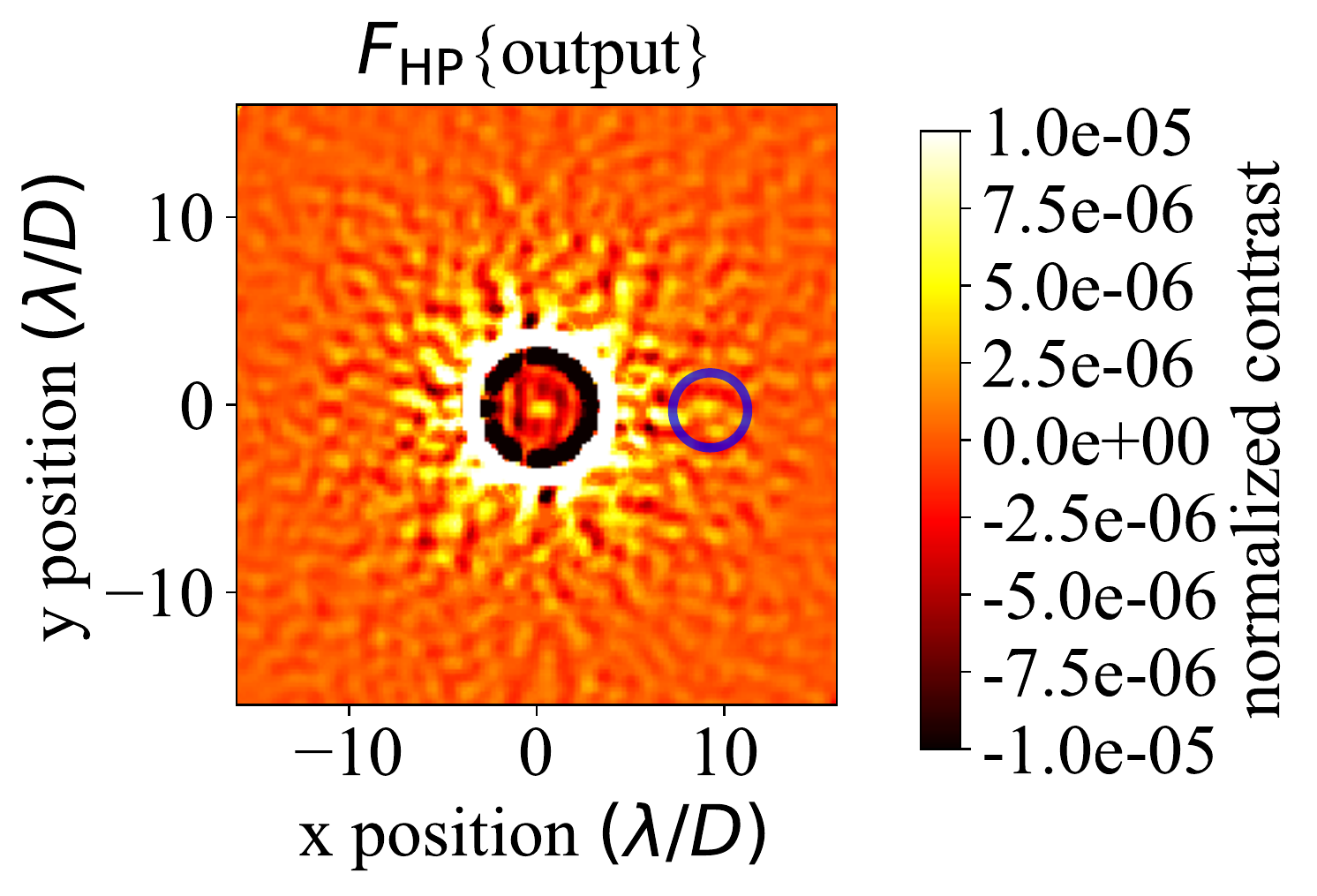}
		(c)
		\label{fig: b}
		\end{center}
	\end{minipage}
	\begin{minipage}[b]{0.45\textwidth}
		\begin{center}
		\includegraphics[width=1.0\textwidth]{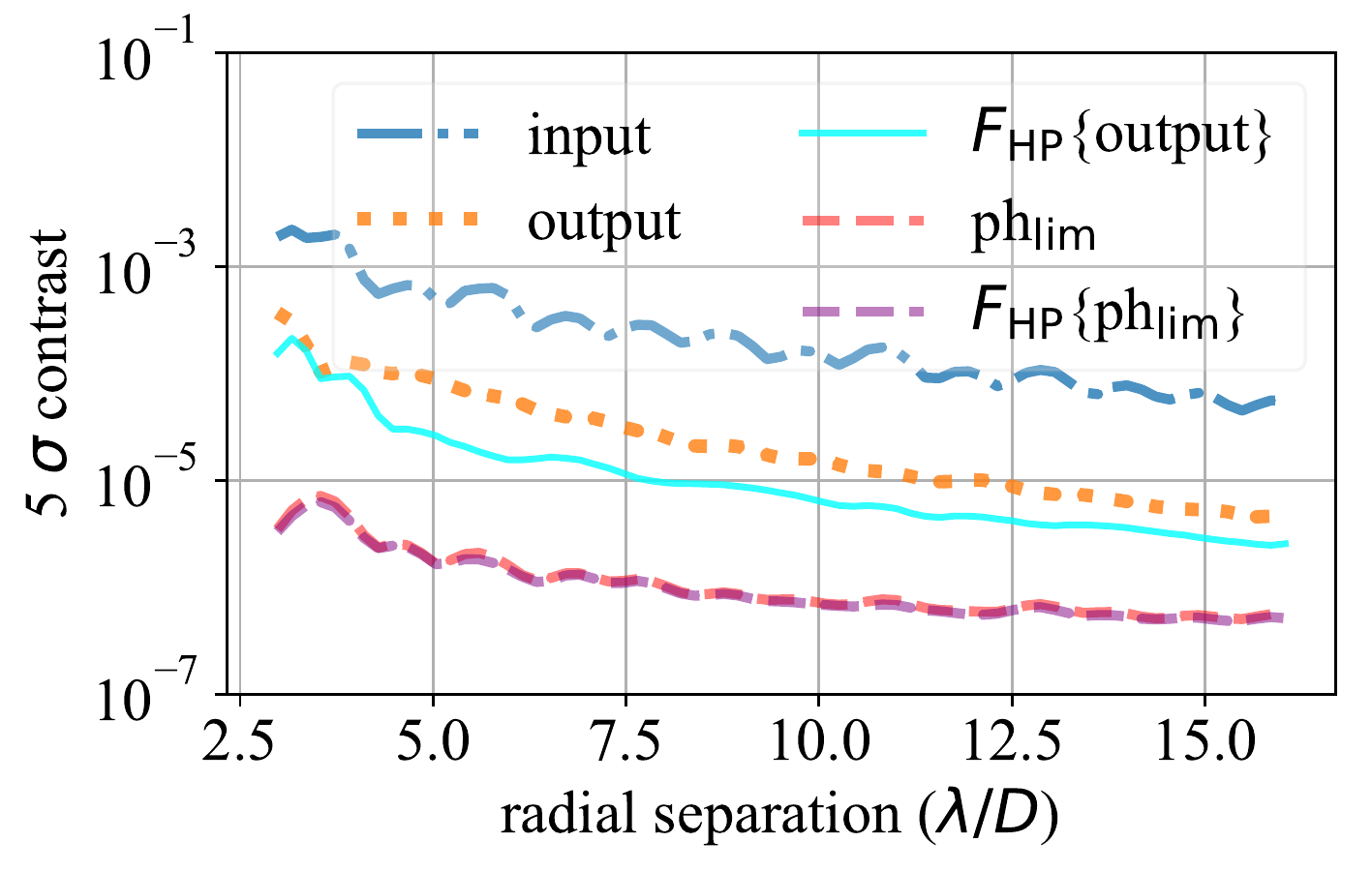}
		(d)
		\label{fig: b}
		\end{center}
	\end{minipage}
	\begin{minipage}[b]{0.45\textwidth}
		\begin{center}
		\includegraphics[width=1.0\textwidth]{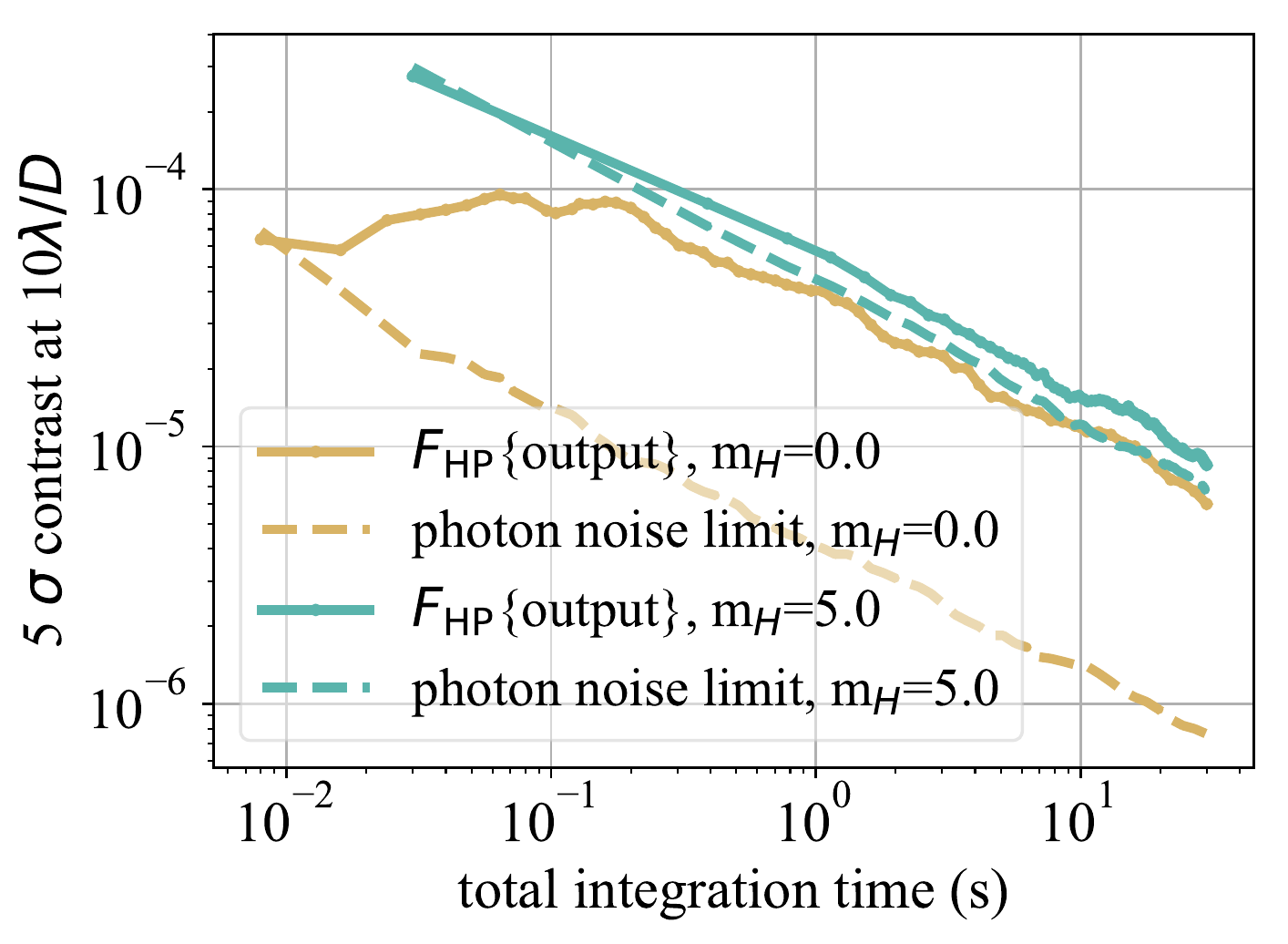}
		(e)
		\label{fig: b}
		\end{center}
	\end{minipage}
\caption{(a) A recorded image for a 30 second exposure of a $m_H=0$ star. (b) A subtraction of the left image using equation \ref{eq: noise_propagation}, assuming a perfect simultaneous measurement of the pinhole PSF during that 30 second exposure. (c) a high-pass filter of (b); a simulated exoplanet at $(x,y)=(10,0)\; \lambda/D$ at a flux-normalized contrast of $10^{-5}$ is identified by the blue circle. (d) a contrast curve of images (a) - (c) compared to the photon noise limit images (labeled ``ph$_\text{lim}$''). (e) Contrast at 10 $\lambda/D$ vs time after applying the same subtraction algorithm as used on image (c) to stacked images at all times in our simulation for both 0th and 5th magnitude stars; the photon noise limit is also shown for comparison to each star.}
\label{fig: long_exp}
\end{figure}

In this section we argue that even with the advent of our new SCC FPM design, the typical CDI strategy of taking long exposures to measure and correct for quasi-static aberration is still limited by the atmosphere. Using the simulation parameters and SCC FPM design presented in appendix \ref{sec: setup} and \S\ref{sec: scc_fpm}, respectively, Figure \ref{fig: long_exp} illustrates the sensitivity limitations for a 30 second exposure of a $m_H=0$ star; it shows (a) the recorded image, (b) a perfect subtraction of everything that is fringed in image (a) (i.e., using equation \ref{eq: noise_propagation}, assuming a noiseless, simultaneous measurement of the pinhole PSF during that 30 second exposure), (c) a high-pass filter of this subtracted image using a $2\times2\;\lambda/D$ median boxcar filter, (d) a contrast curve of these three images compared to the input and photon noise limit images, and (e) the contrast at 10 $\lambda/D$ vs. time of subtracted, high-pass-filtered images compared to the photon noise limit for both $m_H=0$ and $m_H=5$ stars. 

We find that the contrast in Figure \ref{fig: long_exp} (b) is about 50\% of the contrast in the input image (a), a non-negligible component of the image that is `invisible'' to the SCC (or any CDI algorithm) because fringed atmospheric speckles are blurred out over the long exposure. The spatial scales in the residual atmospheric halo of Figure \ref{fig: long_exp}, labeled ``output,'' suggests that a high-pass filter of this image will improve the contrast, and is thus shown to the right as ``$F_\text{HP}$\{output\}.'' Indeed, Figure \ref{fig: long_exp} (d) shows that high-pass filtering does improve the contrast by up to a factor of about 5 throughout the AO control region, although this is still a factor of about 3-20 above the photon noise limit. This contrast curve also illustrates that applying the same high-pass filter to the normal photon noise limit, derived from appendix \ref{sec: phlim}, has a negligible effect.

As discussed in \S\ref{sec: intro}, Figure \ref{fig: long_exp} (e) illustrates the conclusions from \cite{speckle_lifetime}. For a 5th magnitude star, $F_\text{HP}\left\{\text{output}\right\}$ is photon noise-dominated and so contrast continually improves proportional to $t^{-0.5}$. For the 0th magnitude star, $F_\text{HP}\left\{\text{output}\right\}$ is flat and dominated by atmospheric speckles at $t<\tau_\text{spec}\approx200$ ms, and averaging proportional to $(t/\tau_\text{spec})^{-0.5}$ at $t<\tau_\text{spec}$. More generally, we note that although the level of residual atmospheric correlation/speckle lifetime will change with varying observing conditions (causing a variable gap in contrast), the main point of Figure \ref{fig: long_exp} is to illustrate the following averaging properties for atmospheric speckles during long exposures:
\begin{enumerate}
\item Atmospheric speckles leave a residual halo that cannot be measured optically by the SCC or any CDI algorithm.
\item On bright stars, this ultimately limits the achievable contrast, even after post processing, to a non-negligible factor above the photon noise limit.
\item On dim stars, a negligible amount of photons are recorded per speckle lifetime, and so this effect is either at or buried within the photon noise limit.
\end{enumerate}

Points 1-2 above illustrate a fundamental limitation on achievable raw contrast with any CDI algorithm during a long exposure \citep[e.g.,][]{phase_diversty}. Thus, our next FAST approach, discussed below in \S\ref{sec: sim}, is designed to minimize this effect by running fast exposures every few milliseconds, subtracting these recorded images, and then stacking the subtracted residuals to obtain contrast improvement proportional to $t^{-0.5}$.
\subsection{Estimation of the Pinhole PSF}
\label{sec: est_pin_psf}
The main principle of the SCC \citep[][and subsequent papers]{scc_orig, scc_lyot} is summarized in appendix \ref{sec: scc}. From this section and previous work, we note that two different Fourier filtering algorithms on the recorded SCC image, $I$, provide 
\begin{align}
\label{eq: I_minus}\Iminus\equiv |F^{-1}\{F\{I\} \; m_1\}|&= |A_S|\; |A_R|,\text{ and} \\
\label{eq: im}F^{-1}\{F\{I\} \; m_2\} &= |A_S|^2+|A_R|^2+|A_P|^2,
\end{align}
where $m_1$ and $m_2$ are different binary masks, and $A_S,\; A_R,\; \text{and}\; A_P$ are the electric field in the detector plane from the stellar speckles, reference pinhole, and exoplanet, respectively. 

From these two equations with three unknows, we cannot retrieve the exoplanet image $|A_P|^2$. \citet{baudoz_psfsubt} approach this problem by recording a calibration image of $I_R\equiv|A_R|^2$ (i.e., the pinhole PSF), using an internal source and occulting the main beam of the Lyot stop. This is a viable approach to reconstructing the speckle pattern $|A_S|^2$ induced by small and/or static aberrations. However, it may not be for measuring atmospheric speckles in millisecond timescale exposures unless the electric field going through the pinhole is relatively stable across multiple realizations of a translating atmospheric phase screen. We discuss this calibrated reference imaging approach later in \S\ref{sec: cal}. In \S\ref{sec: filt}, we discuss the approach of direct, on-sky measurement of the pinhole PSF. Finally, in \S\ref{sec: algo} we present a filtering algorithm, using only the recorded on-sky SCC image to reconstruct the pinhole PSF.
\subsubsection{Direct Measurement}
\label{sec: filt}
Obtaining a high, simultaneous SNR image of the pinhole PSF along with a noisy target image should result in a contrast improvement proportional to $t^{-0.5}$ (see appendix \ref{sec: post_processing} and \ref{sec: phlim} for a further discussion). However, in reality this measurement of the pinhole PSF will also be affected by photon noise due to the finite amount of light transmitted through the SCC pinhole. Using the SCC FPM, we simulate a separate pinhole PSF measurement assuming use of a 50/50 beam splitter placed on the pinhole: 50\% of the light going through the pinhole is transmitted through the beam splitter and focused onto the science camera, while the other 50 \% is redirected by a small angle (i.e., just downstream of the pinhole) and focused onto the same camera.  This would also require a field stop in an upstream focal plane (either at or downstream from the FPM plane) to block any stellar background that would otherwise prevent any pinhole PSF light from being recorded with a reasonable SNR. The beam splitter will induce quasi-static and chromatic aberrations on the reference beam, but we do not simulate them in this paper because their effects should be second order as the beam splitter is downstream from the FPM and can be super-polished to extremely high level precision.

\begin{figure}[!h]
	\begin{minipage}[b]{0.35\textwidth}
		\begin{center}
		\includegraphics[width=1.0\textwidth]{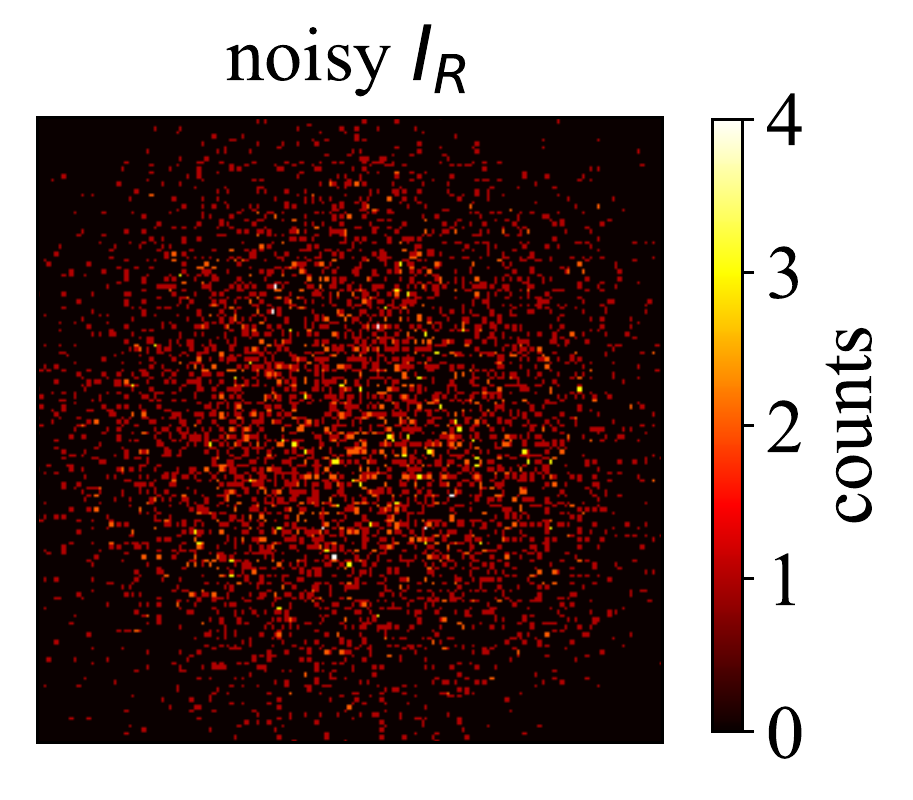}
		(a)
		\label{fig: a}
		\end{center}
	\end{minipage}
	\begin{minipage}[b]{0.28\textwidth}
		\begin{center}
		\includegraphics[width=1.0\textwidth]{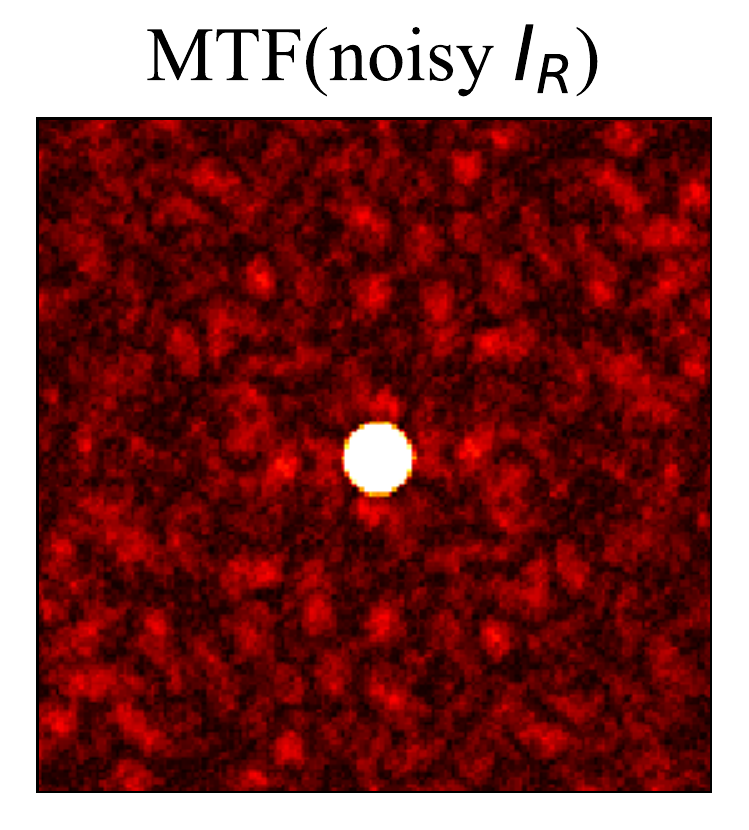}
		(b)
		\label{fig: b}
		\end{center}
	\end{minipage}	
	\begin{minipage}[b]{0.28\textwidth}
		\begin{center}
		\includegraphics[width=1.0\textwidth]{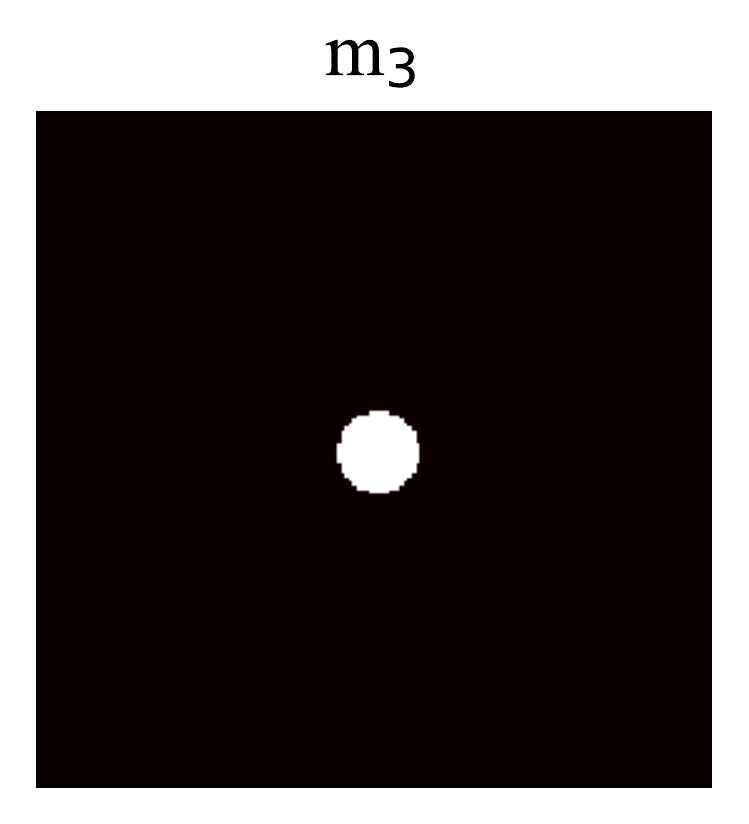}
		(c)
		\label{fig: c}
		\end{center}
	\end{minipage}	
	\begin{minipage}[b]{0.3\textwidth}
		\begin{center}
		\includegraphics[width=1.0\textwidth]{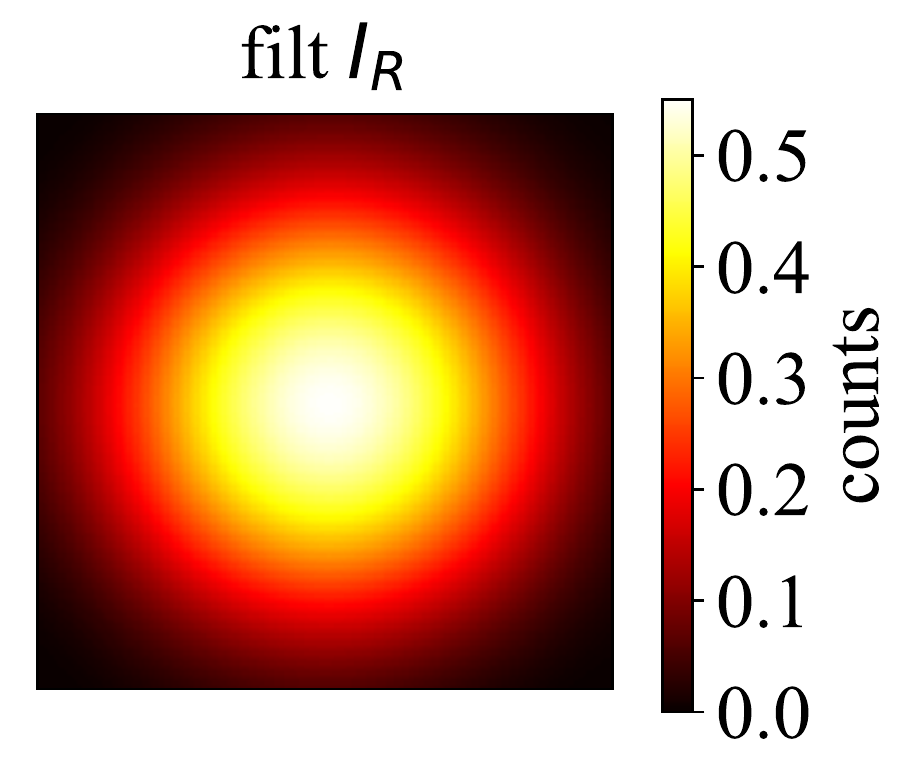}
		(d)
		\label{fig: d}
		\end{center}
	\end{minipage}	
	\begin{minipage}[b]{0.3\textwidth}
		\begin{center}
		\includegraphics[width=1.0\textwidth]{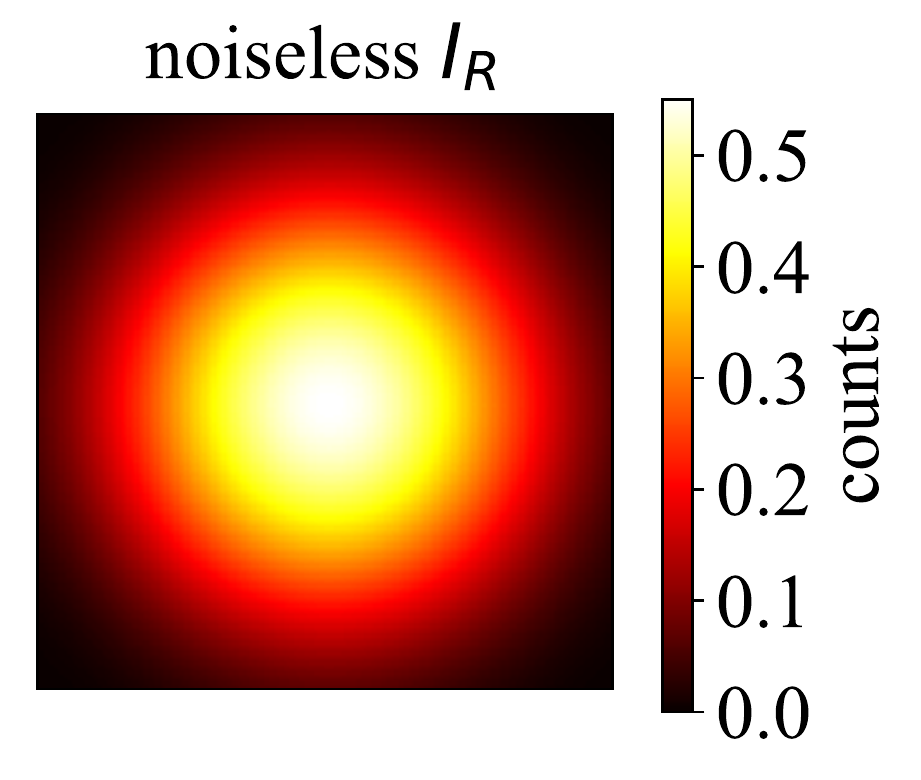}
		(e)
		\label{fig: d}
		\end{center}
	\end{minipage}	
	\begin{minipage}[b]{0.33\textwidth}
		\begin{center}
		\includegraphics[width=1.0\textwidth]{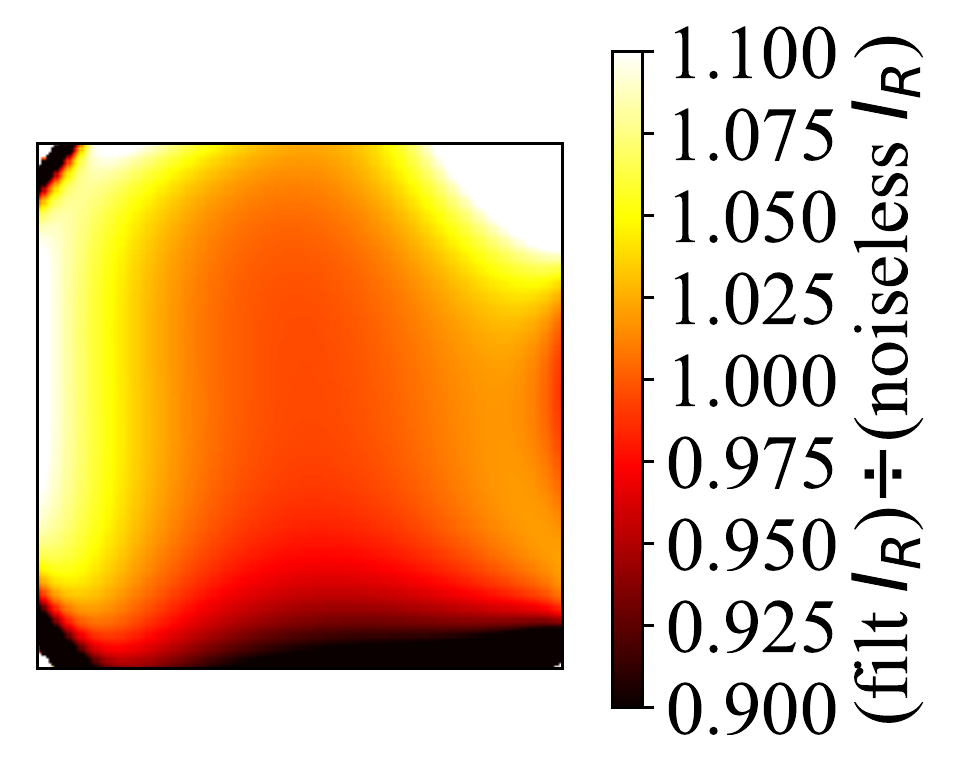}
		(f)
		\label{fig: f}
		\end{center}
	\end{minipage}	
\caption{(a) a recorded pinhole PSF, with photon noise, for a 1 ms integration on a m$_H$=0 star using the SCC FPM. (b) the MTF of (a). (c) a binary mask used to Fourier filter (a) so that photon noise is suppressed on spatial-scales below $\lambda/d$. (d) the result of the result of Fourer filtering (a): filt $I_R=F^{-1}\{F\{\text{noisy I}_R\} m_3\}$. (e) a recorded pinhole PSF where no photon noise is simulated. (f) the ratio of (d) to (e).}
\label{fig: imr}
\end{figure}

With the setup discussed above, the pinhole PSF for a one millisecond exposure of a $m_H$=0 star is shown in Figure \ref{fig: imr}(a). Although this image is indeed recording as much as four photons per pixel in a one millisecond exposure, photon noise still clearly dominates compared to the image without photon noise in \ref{fig: imr}(e). However, looking at the MTF of \ref{fig: imr}(a) in \ref{fig: imr}(b) shows that only the central region of Fourier space has physical representation of spatial-scales larger than $\lambda/d$ (i.e., the central region of the MTF), while all other higher spatial frequencies in the MTF are only from recorded photon noise. Thus, using the Fourier filter $m_3$ in \ref{fig: imr}(c) allows the noisy image in \ref{fig: imr}(a) to be converted to the Fourier-filtered image in \ref{fig: imr}(d). The ratio of \ref{fig: imr}(d) to \ref{fig: imr}(e), illustrated in \ref{fig: imr}(f), shows that this reconstruction, even for an extremely low number of recorded photons is still accurate to about $\pm$10\%. As long as this offset is random on a frame by frame basis (i.e. generated from photon noise), stacking the output of a subtracted image using equations \ref{eq: I_minus} and \ref{eq: im} (i.e., where a filtering algorithm as in \ref{fig: imr}(d) is used to estimate the pinhole PSF) will continue to average randomly (see appendix \ref{sec: phlim} for a further discussion). Only a static offset generating speckles that are consistently over- or under-subtracted will create static residuals after stacking. This same principle applies to the other two pinhole PSF reconstruction methods in \S\ref{sec: cal} and \S\ref{sec: algo}. Intuitively, there is no reason to believe here that any static offset would be present in Figure \ref{fig: imr}(e), since these residuals are generated from photon noise present at the same low spatial frequencies as the pinhole PSF, and that low-order diffraction and/or atmospheric effects from the FPM are unlikely to cause such static offsets (but see \S\ref{sec: fluxnorm}). 
\subsubsection{Calibration Algorithm}
\label{sec: cal}
Similar to the approach in \citet{baudoz_psfsubt}, a daytime pinhole PSF can be recorded with very high SNR using an internal source and masking the main beam in the Lyot plane. However, this recorded calibration pinhole PSF will be without
\begin{enumerate}
\item an intrinsic flux normalization, unlike the direct measurement in \S\ref{sec: filt}, and
\item effects from residual AO atmospheric turbulence
\end{enumerate}
The implications of point 2 above are that additional low order aberrations from the atmospheric phase screen (i.e., at spatial frequencies of less than 3 cycles/pupil for a 6 $\lambda/D$ diameter SCC FPM) that are not present in the calibration pinhole PSF will ultimately limit the achievable contrast with this method. This limitation arises from non-centro-symmetric features of the on-sky pinhole PSF relative to the static pinhole PSF (discussed further in appendix \ref{sec: fluxnorm}). Furthermore, we simulate the static phase screen differences between a daytime calibration and on-sky measurement by using a different specific wavefront map of static phase and amplitude aberration for the calibration pinhole PSF vs. the on-sky static component, where both are still normalized to the same rms values as described in appendix \ref{sec: setup} but are fully decorrelated from one another. Then, assuming we have properly calibrated the pinhole PSF with this approach, equations \ref{eq: I_minus} and \ref{eq: im} allow a subtraction of $|A_S|^2$ and $|A_R|^2$ without subtracting $|A_P|^2$.

\subsubsection{Reconstruction Algorithm}
\label{sec: algo}
In this section we will revisit the underdetermined system of equations \ref{eq: I_minus} and \ref{eq: im} (i.e., two equations and three unknowns). If $|A_P|^2$ could be removed from equation \ref{eq: im}, we would have a determined system. Noting that the spatial scale of $|A_R|^2$ is limited to $\lambda/d$, running a low-pass filter on equations \ref{eq: I_minus} and \ref{eq: im} to retain spatial scales larger than or equal to $\lambda/d$ should have the effect of ``filtering the exoplanet light out of the image," thus providing two equations with two unknown and a solution for $|A_R|$. These steps are outlined below, where $\lpf\{\}$ is a linear low-pass filter operator: 
\begin{align}
\Iminuslpf \equiv \lpf\{\Iminus\} &= \lpf\{|A_S|\; |A_R|\} \nonumber \\
\label{eq: Iminuslpf} &\approx\lpf\{|A_S|\}|A_R|, \text{ and} \\
\imlpf \equiv \lpf\{F^{-1}\{F\{I\} m_2\}\} &= \lpf\{|A_S|^2+|A_R|^2+|A_P|^2\} \nonumber \\
&\approx \left[\lpf\{|A_S|\}\right]^2+\left[\lpf\{|A_P|\}\right]^2+|A_R|^2 \nonumber\\
\label{eq: imlpf} &\approx \left[\lpf\{|A_S|\}\right]^2+|A_R|^2.
\end{align}
We found that using a 3.9$\times$3.9 $\lambda/D$ median boxcar filter for $\lpf\{\}$ provided both a good contrast and algorithmic exoplanet throughput (see appendix \ref{sec: setup} for our formal definition), although these metrics were not optimized with any tolerance requirements. 

We are able to remove the $\lpf\{\}$ from $|A_R|$ because a 3.9 $\lambda/D$ median boxcar kernel has a negligible impact on the smallest $\lambda/d$ scales ($\lambda/d=37\; \lambda/D$ in our simulation) of $|A_R|$. Additionally, in equation \ref{eq: imlpf} we assume that $\left[\lpf\{|A_P|\}\right]^2\approx0$ because the central $\lambda/D$-sized core of the exoplanet PSF is removed by the low pass filter (only about 2\% of the peak flux in an exoplanet core remains after applying the low pass filter), and we ignore the higher order halo terms of the exoplanet PSF in the context of point source detection \citep{perrin_psf}. However, we do not make the same argument for the low-pass filtered starlight term, $\lpf\{|A_S|^2\}$; because this term is of similar magnitude to the $|A_R|^2$ term, ignoring the low spatial frequency starlight information would otherwise bias our estimate of $|A_R|^2$. Ultimately, the exoplanet PSF halo term is absorbed into $\lpf\{|A_S|^2\}$ and/or $|A_R|^2$, which will eventually cause static biases in the high exoplanet SNR regime. Thus, combining equations \ref{eq: Iminuslpf} and \ref{eq: imlpf} (now two equations and two unknowns), we find that
\begin{equation}
|A_R|^2=\frac{\left(\imlpf-\sqrt{\imlpf^2-4\; \Iminuslpf^2}\right) }{2} .
\label{eq: Ar}
\end{equation}
Finally, combining equations \ref{eq: I_minus}, \ref{eq: im}, and \ref{eq: Ar} then allows us to subtract all of the terms in equation \ref{eq: im} except for $|A_P|^2$.
\subsection{Flux Normalization and Static Limitations}
\label{sec: fluxnorm}
In this section we present a flux normalization procedure and discuss static limitations of the subtraction algorithms presented in \S\ref{sec: est_pin_psf}. For all algorithms, we found additional static systematic offsets that ultimately prevented contrast improvement below a certain level. For this reason, we implemented two additional steps to flux normalize the reconstructed image without subtracting any additional exoplanet flux:
\begin{enumerate}
\item annular flux normalization by summation:
\begin{align}
\text{im}_\text{subt}(r)&=F^{-1}\{\ F\{\text{I}_\text{noisy}\}m_2\}-c_\text{norm}(r)\left[ |F^{-1}\{F\{\text{I}_\text{noisy}\}m_1\}|^2/\tilde{I}_R+\tilde{I}_R \right] ,\text{ where} \nonumber \\
\text{c}_\text{norm}(r)&=\sum_r\left[F^{-1}\{\ F\{\text{I}_\text{noisy}\}m_2\}\right]/\sum_r\left[|F^{-1}\{F\{\text{I}_\text{noisy}\}m_1\}|^2/\tilde{I}_R+\tilde{I}_R\right],
\label{eq: norm}
\end{align}
where $\text{I}_\text{noisy}$ represents the recorded noisy SCC image, $\tilde{I}_R$ represents the estimated/calibrated pinhole PSF via any of the algorithms presented so far, and $\sum\limits_r$ denotes a sum of all the pixels within 0.3 $\lambda/D$ of a given separation $r$. Equation \ref{eq: norm} is equivalent to forcing the mean of the output subtracted image, im$_\text{subt}$, to 0. To note, even though the target image, $F^{-1}\{\ F\{\text{I}_\text{noisy}\}m_2\}$, includes exoplanet light and therefore biases the normalization coefficient at the separation of the exoplanet, we found about 0-15\% algorithmic exoplanet throughput losses from this effect for our simulated exoplanets at flux-normalized contrasts levels of $10^{-5}$ and $5\times10^{-5}$ (see \S\ref{sec: sim}), and otherwise found significant improvement of normalization by summation vs., e.g., robust standard deviation. 

\item High-pass filtering the already-flux-normalized image from equation \ref{eq: norm}, but only after stacking images to the desired integration time. The annular normalization approach in step 1 will not remove symmetric, low-order residual aberrations that can build up over time because of small calibration errors, e.g., as discussed in \S\ref{sec: cal}. Thus, we use an aggressive 2.1 $\times$ 2.1 $\lambda/D$ median boxcar filter to generate a low-pass-filtered version of the flux-normalized subtracted image, $\lpf\left\{\text{im}_\text{subt}\right\}$, and then high-pass filter the image via:
\begin{equation}
F_\text{HP}\left\{\text{im}_\text{subt}\right\}=\text{im}_\text{subt} - F_\text{LP}\left\{\text{im}_\text{subt}\right\}
\label{eq: hpf}
\end{equation}
\end{enumerate}
Equation \ref{eq: hpf} is the final form of stacked images that are displayed and shown in contrast curves of this paper for the calibration and reconstruction algorithms. We found that annular normalization degraded contrast in the case of the direct pinhole PSF measurement algorithm, and so for this algorithm we only use a high-pass filter. However, interestingly, we did find static limitations for the direct pinhole PSF measurement algorithm if we did not use a high-pass filter, revealing the presence of static low order aberrations as a result of our Fourier filtering algorithm illustrated in Figure \ref{fig: imr}. For an analagous comparison we also subsequently high-pass filter the photon noise limit images (see appendix \ref{sec: phlim} but also Figure \ref{fig: long_exp} which shows this has a negligible effect). 

\begin{figure}[!h]
\centering
	\begin{minipage}[b]{0.45\textwidth}
		\begin{center}
		\includegraphics[width=1.0\textwidth]{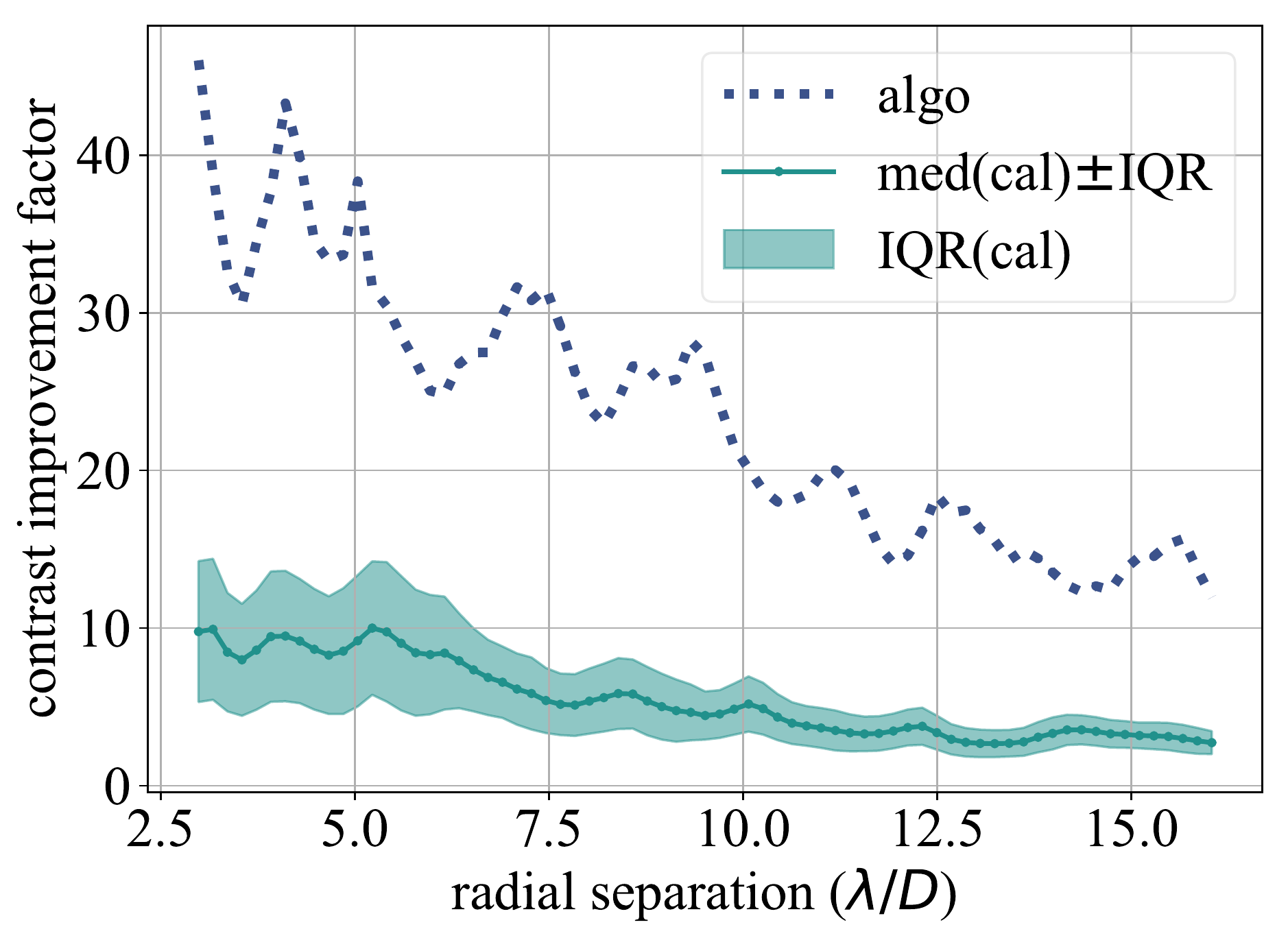}
		(a)
		\label{fig: a}
		\end{center}
	\end{minipage}
	\begin{minipage}[b]{0.45\textwidth}
		\includegraphics[width=1.0\textwidth]{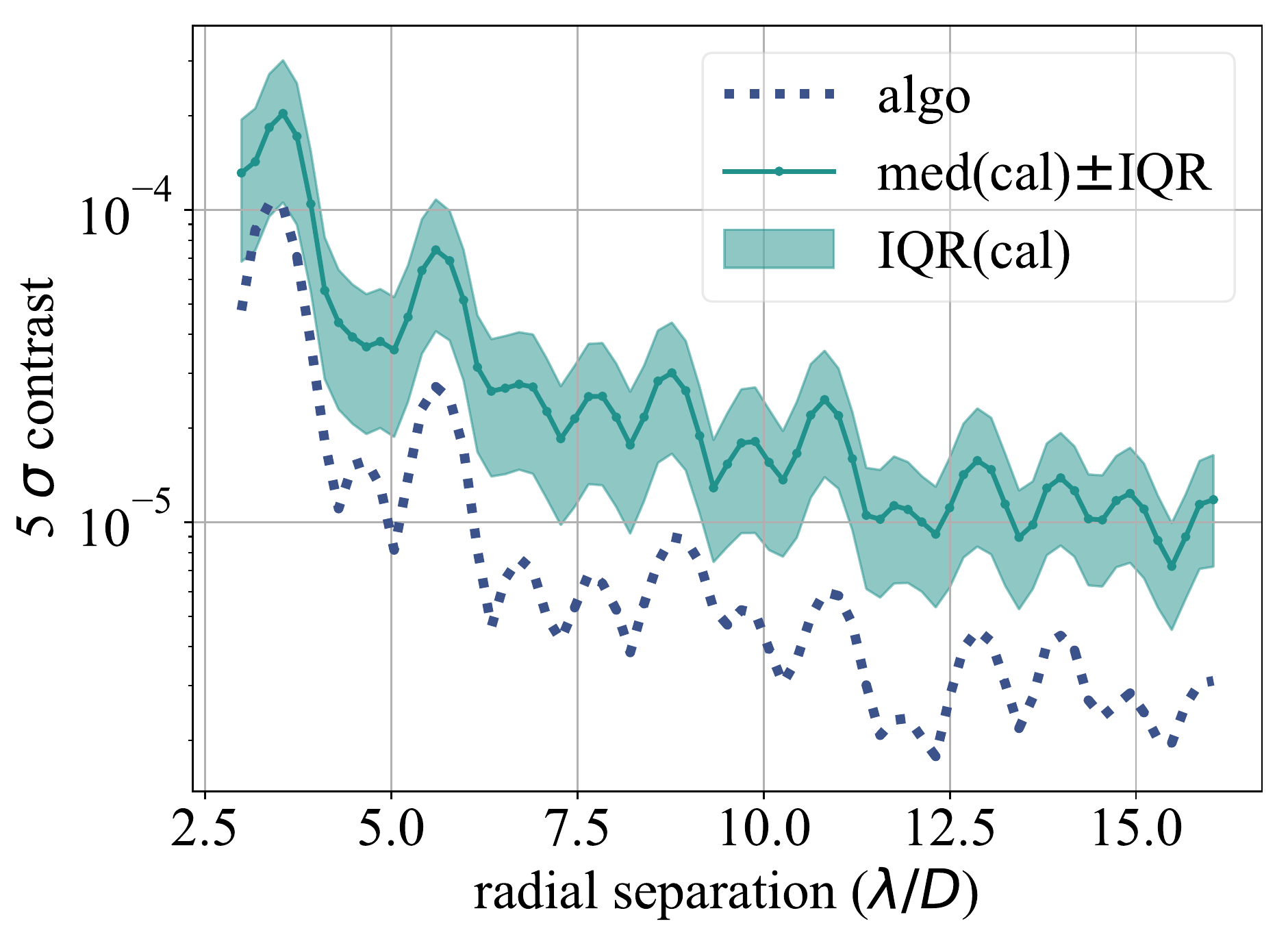}
		\begin{center}
		(b)
		\label{fig: b}
		\end{center}
	\end{minipage}	
	\begin{minipage}[b]{0.37\textwidth}
		\begin{center}
		\includegraphics[width=1.0\textwidth]{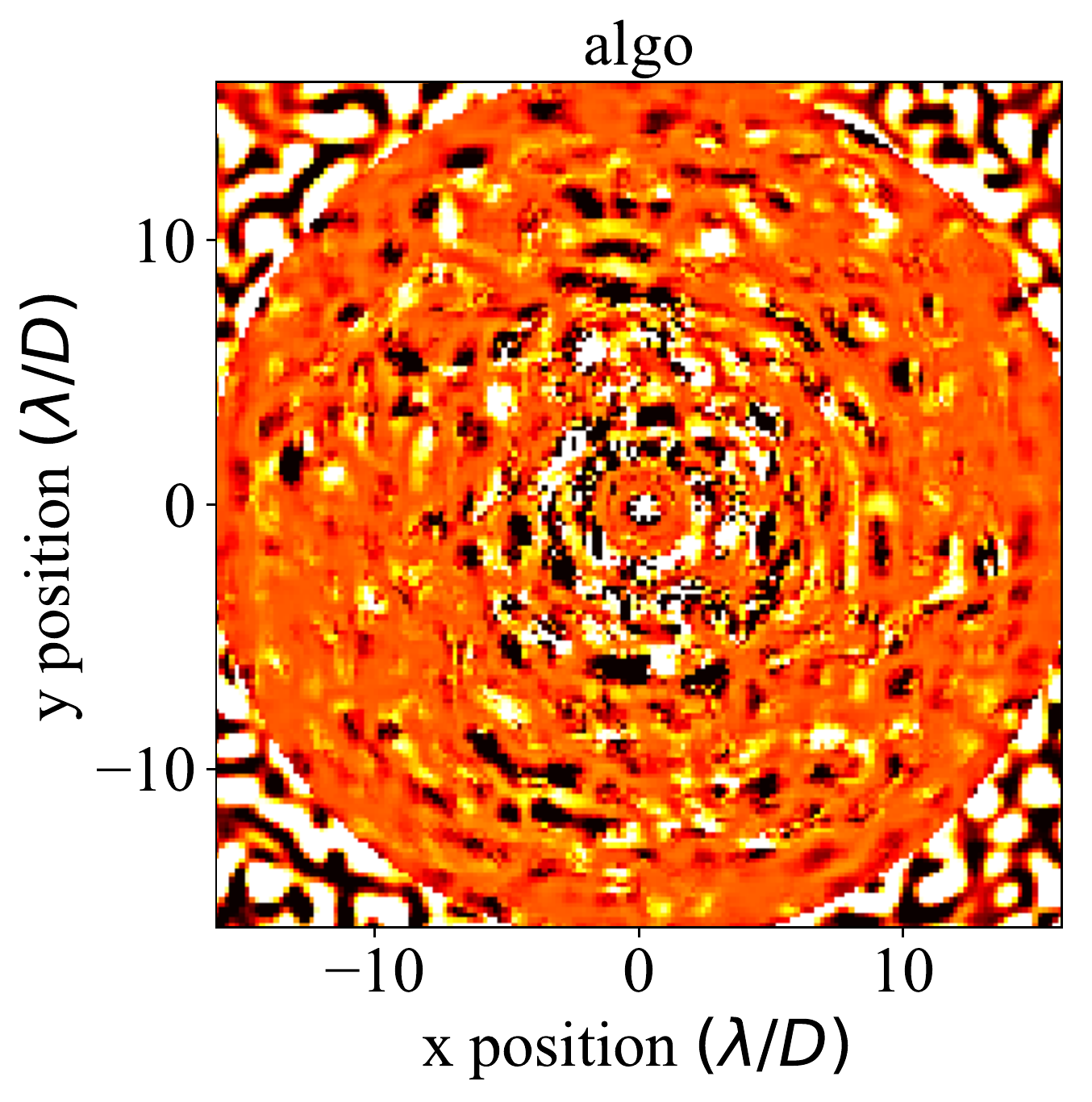}
		(c)
		\label{fig: c}
		\end{center}
	\end{minipage}	
	\begin{minipage}[b]{0.45\textwidth}
		\begin{center}
		\includegraphics[width=1.0\textwidth]{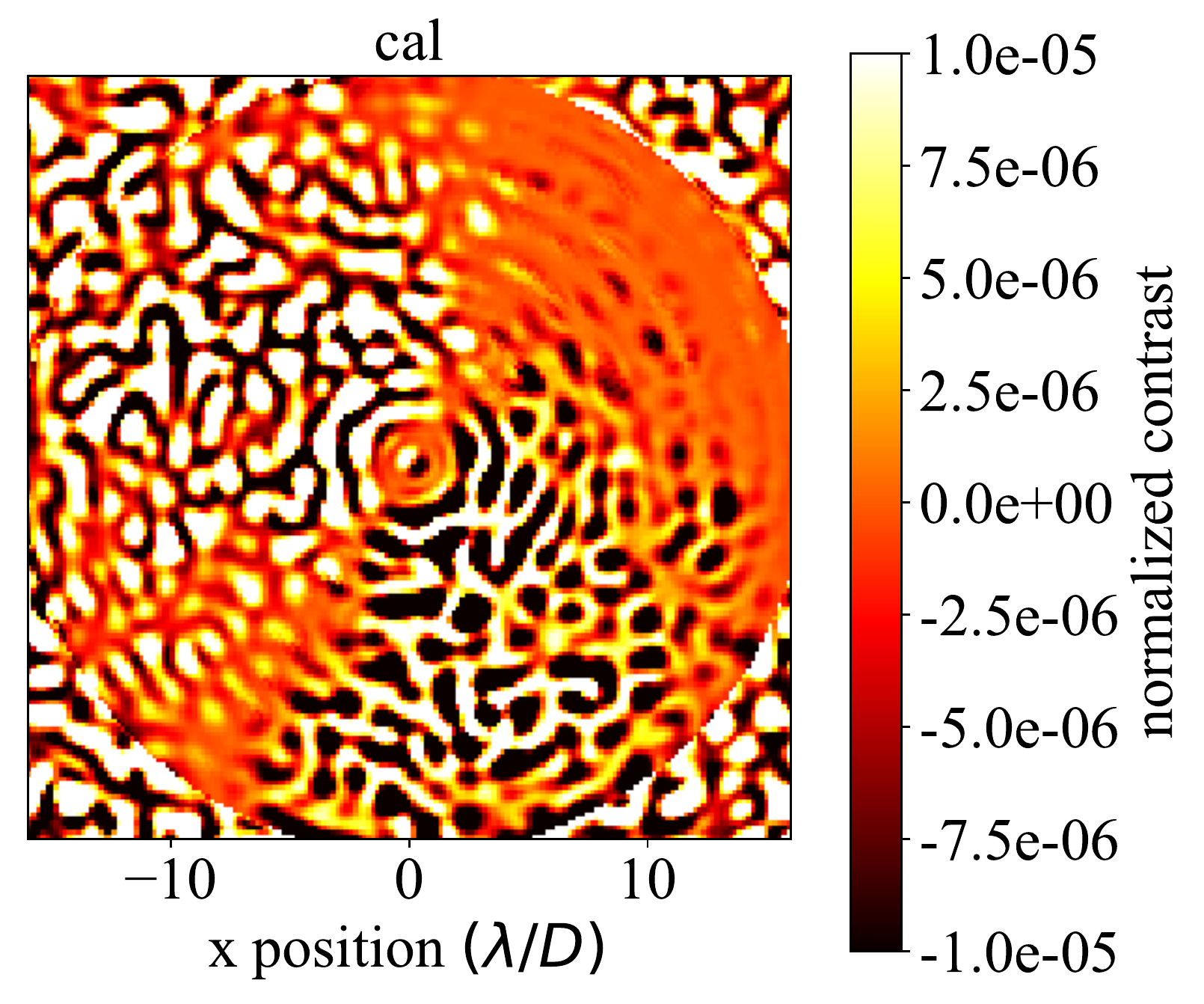}
		(d)
		\label{fig: d}
		\end{center}
	\end{minipage}	
\caption{Performance limitations of the calibration (shown as ``cal'') and pinhole PSF reconstruction (``algo'') algorithms, in the absence of photon noise. (a) contrast improvement factor vs. radial separation for the algo and cal algorithms (i.e., a ratio of contrast curves between the input target image and output subtracted image). The target image includes both atmospheric and static wavefront components (see appendix \ref{sec: setup} for additional details). The cal algorithm is shown as the median and interquartile range (IQR) from 100 different iterations that each use a different static phase screen to generate the calibration pinhole PSF (but still with the same rms amplitude). (b) contrast curves for the algo and cal algorithms applied to a SCC image with only static aberration, predicting the ultimate limits for either algorithm during a long exposure. (c) and (d): the algo and cal images, respectively, used to generate the contrast improvement curves in (a); a single random realization is selected from the 100 different cal iterations.}
\label{fig: clim}
\end{figure}

Even in the absence of photon noise, there is still a fundamental limit on the performance of the calibration and reconstruction algorithms, shown in Figure \ref{fig: clim}. The labels ``algo,'' ``cal,'' and ``filt'' represent the reconstruction, calibration, and direct pinhole PSF algorithms presented in sections \ref{sec: algo}, \ref{sec: cal}, and \ref{sec: filt}, respectively. Without photon noise, the direct pinhole PSF is limited by numerical noise at the $10^{-18}$ level (see appendix \ref{sec: phlim}), and so this subtraction algorithm does not suffer from the same level of performance degradation in terms of achievable contrast improvement like the other two do.

Figure \ref{fig: clim} (a) shows the factor by which contrast improves compared to an input noiseless target image (but still with atmospheric and static wavefront components as described in appendix \ref{sec: setup}) and thus illustrates that we never expect to improve contrasts in a single subtraction by better than a factor of about 20 and 6 at 10 $\lambda/D$ with the algo and cal algorithms, respectively. Because no photon noise is simulated here, these limits are related to imperfections in the estimate of the pinhole PSF, which ultimately cause an asymmetric speckle pattern in the image that biases the normalization coefficient from equation \ref{eq: norm}. Specifically, this bias arises from the $\left[ |F^{-1}\{F\{\text{im}_\text{noisy}\}m_1\}|^2/\tilde{I}_R+\tilde{I}_R \right]$ term generated for the pinhole PSF calibration/reconstruction, since a combination of phase and amplitude aberration in the pupil plane produces a non-centro-symmetric speckle pattern (see appendix \ref{sec: setup} for a further discussion). This asymmetry will limit the achievable contrast in equation \ref{eq: norm} by using a single normalization coefficient at a fixed separation. However, over a long exposure the limits from Figure \ref{fig: clim} (a) may not be the same; a large number of uncorrelated residual atmospheric speckle patterns may average to allow reaching greater improvement than the limits for a single atmospheric phase screen realization.

The predicted contrast limits for a long exposure are illustrated in Figure \ref{fig: clim} (b), showing the subtraction of a target image with only static aberration (i.e., no atmospheric or photon noise component is simulated). The atmospheric averaging effect described above should in principle not reach below the limits in Figure \ref{fig: clim} (b). Thus, once adding photon noise, which should in principle not create additional static effects, the contrast improvement proportional to $t^{-0.5}$ from stacking images is expected to flatten out around these levels.

Figure \ref{fig: clim} also shows a variation in contrast improvement for the calibration algorithm depending on the specific phase screen realization used to generate the pinhole PSF. This illustrates the sensitivity of this technique to pupil modes of any spatial frequency that vary in each phase screen realization (although tip and tilt are removed by a least-squares from each static phase screen; see appendix \ref{sec: setup} for further details) and/or low-order modes diffracted from the FPM into the off-axis Lyot stop pinhole; one instance of this effect is clearly seen in Figure \ref{fig: clim} (d), which shows residual low order features that span the full AO control region (i.e., every individual subtracted cal image shows a different residual low order pattern). This sensitivity can be either rewarding or degrading, varying contrast factors of $\pm$10 in different parts of the image, and thus warrants a further study of how to favorably optimize this variation (e.g., using a deconvolution procedure to remove the dominant modes that are causing the contrast degradation), but is beyond the scope of this paper.
\section{FAST Simulations and Analysis}
\label{sec: sim}
\begin{figure}[!h]
\centering
	\begin{minipage}[b]{0.4\textwidth}
		\begin{center}
		\includegraphics[width=1.0\textwidth]{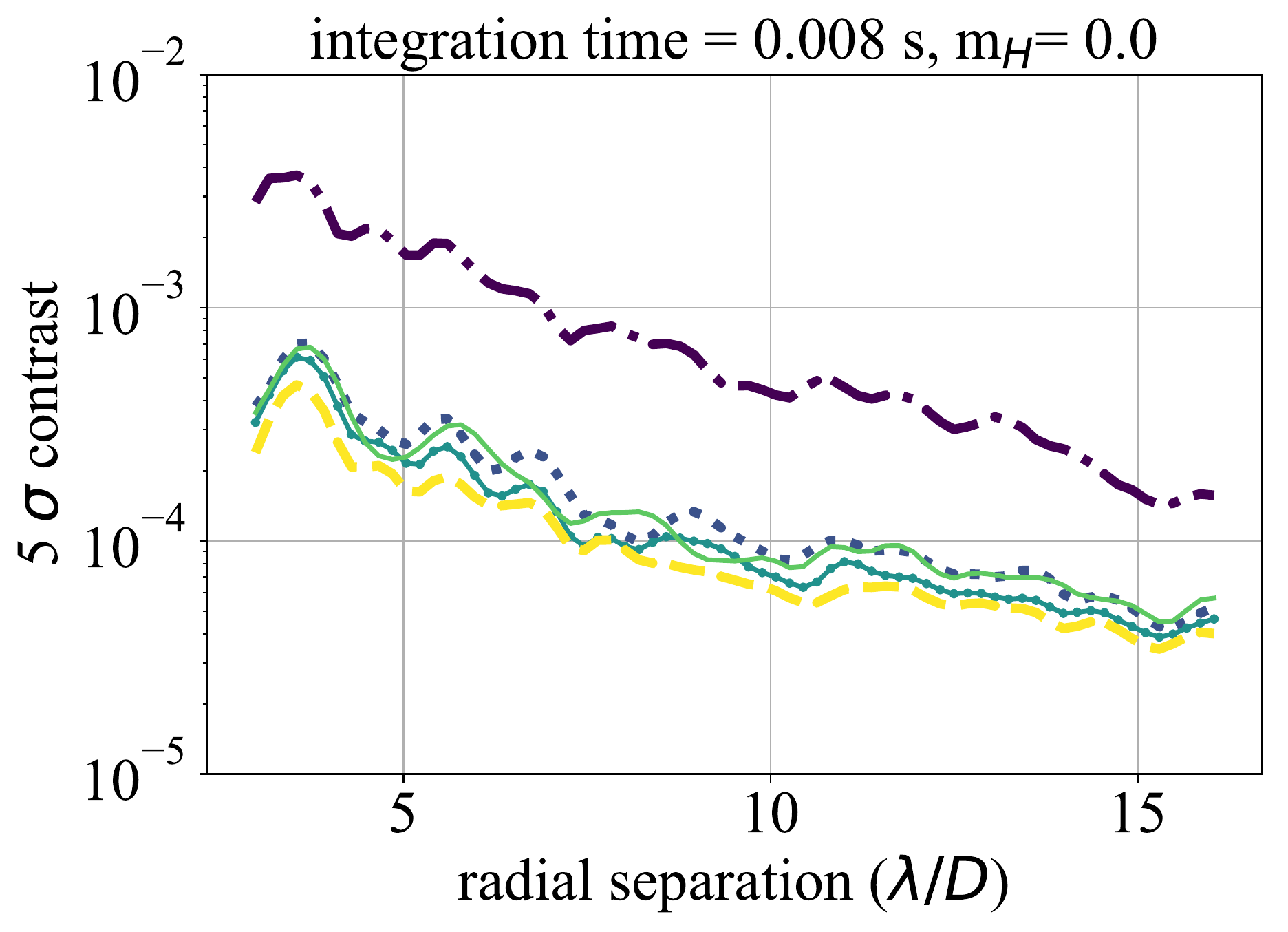}
		(a)
		\label{fig: a}
		\end{center}
	\end{minipage}
	\begin{minipage}[b]{0.4\textwidth}
		\begin{center}
		\includegraphics[width=1.0\textwidth]{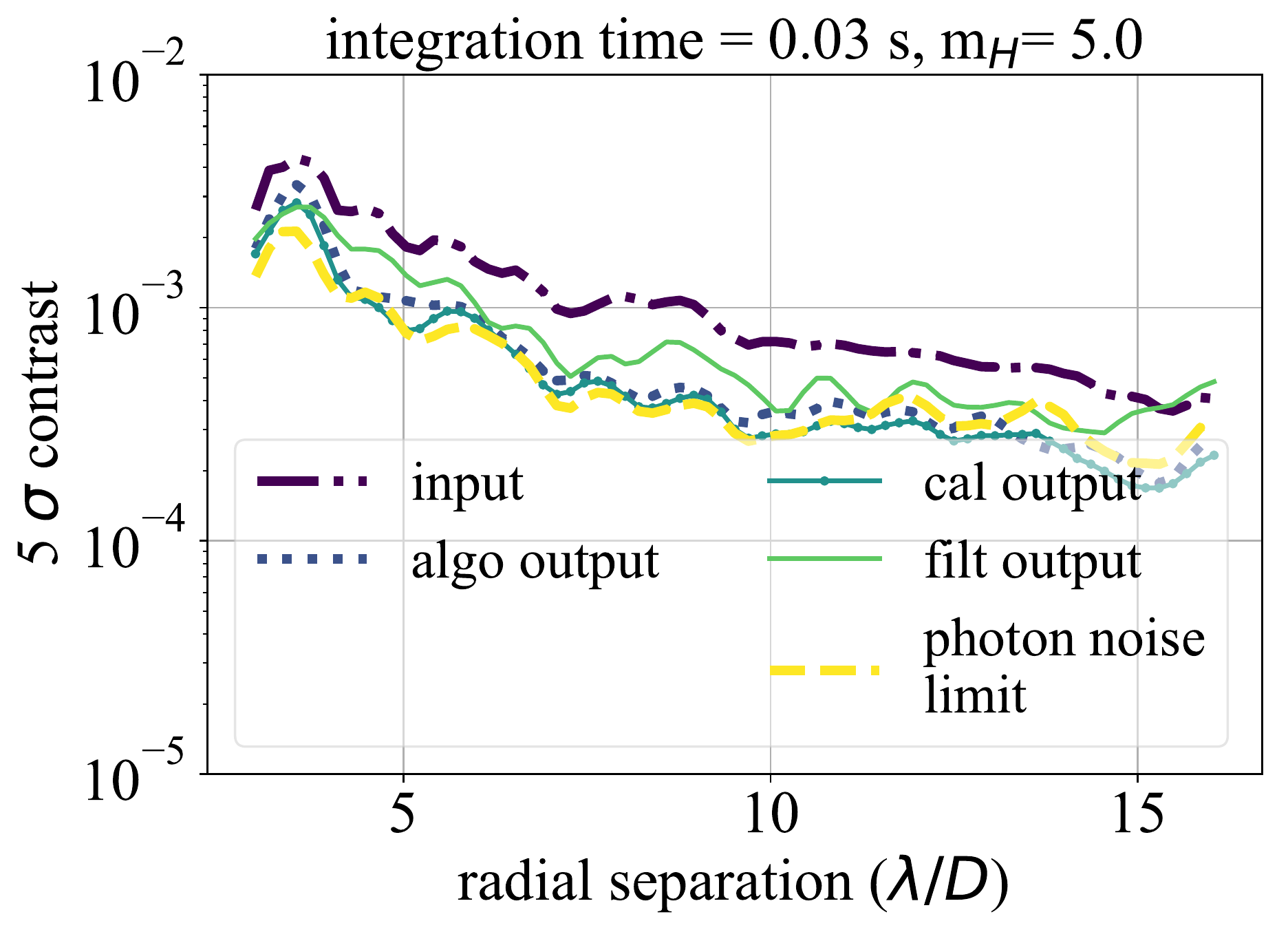}
		(b)
		\label{fig: b}
		\end{center}
	\end{minipage}	
\caption{Contrast curves showing the recorded input image, output subtracted images, and photon noise limit for (a) an 8 millisecond exposure on a 0th magnitude star, and (b) a 30 millisecond exposure on a 5th magnitude star. The legend in (b) is the same for all panels. The labels ``algo,'' ``cal,'' and ``filt'' represent the reconstruction, calibration, and direct pinhole PSF algorithms presented in sections \ref{sec: algo}, \ref{sec: cal}, and \ref{sec: filt}, respectively.}
\label{fig: short_exp}
\end{figure}

With the framework developed in \S\ref{sec: fast}, we consider the FAST approach to measuring and subtracting both quasi-static and atmospheric speckles over short exposures, in contrast to the long exposure strategy in \S\ref{sec: long_exp}. Figure \ref{fig: short_exp} shows the results of recording and subtracting a 8 millisecond exposure of a 0th magnitude star and 30 millisecond exposure of a 5th magnitude star according to each method described in \S\ref{sec: est_pin_psf}. 
Figure \ref{fig: short_exp} illustrates that we are able to reach close to the photon noise limit in a relatively short exposure for both a 0th and 5th magnitude star, although this limit is only a factor of about 10 times below the input target image. Assuming the use of a photon counting camera, photon noise will dominate in these short exposures. As already discussed in \S\ref{sec: intro} and illustrated in Figure \ref{fig: long_exp} (e), the speckle lifetime in our simulations is about 200 milliseconds. Thus, the choice of a 30 millisecond integration time for the $m_H=5$ simulations does not risk the AO halo averaging effect over multiple speckle lifetimes. With this concept in mind of running fast enough to ``freeze'' the atmospheric speckles, we also choose to use 8 ms exposures on the 0th magnitude simulation. Although a 1 kHz frame rate also reaches the same photon noise limit after 30 seconds, the former 125 Hz frame rate provides a larger contrast improvement in individual subtractions, as illustrated in the left vs. right panels of Figure \ref{fig: short_exp}.

\begin{figure}[!h]
\begin{center}
	\begin{minipage}[b]{0.45\textwidth}
		\includegraphics[width=1.0\textwidth]{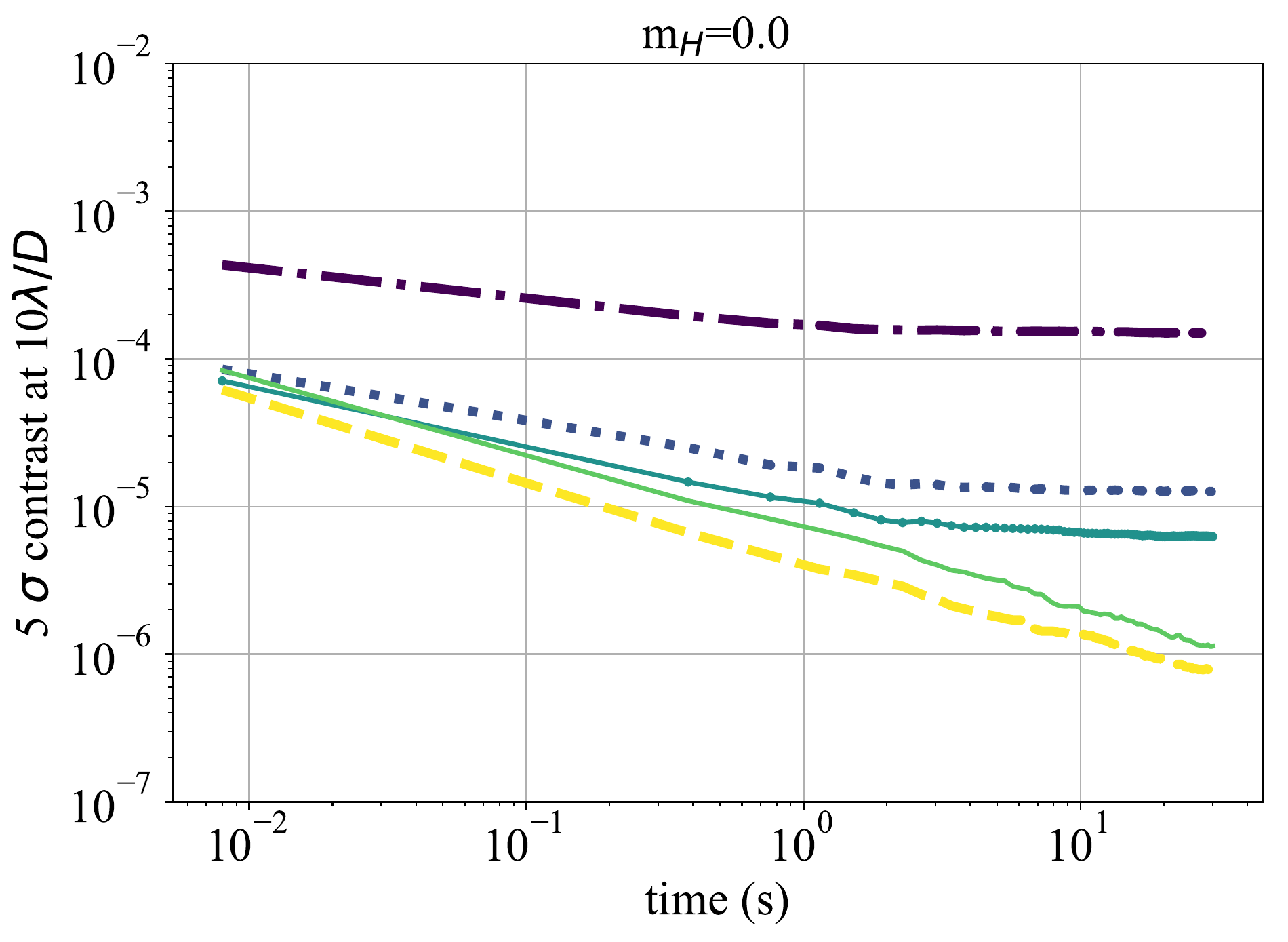}
		\label{fig: a}
	\end{minipage}
	\begin{minipage}[b]{0.45\textwidth}
		\includegraphics[width=1.0\textwidth]{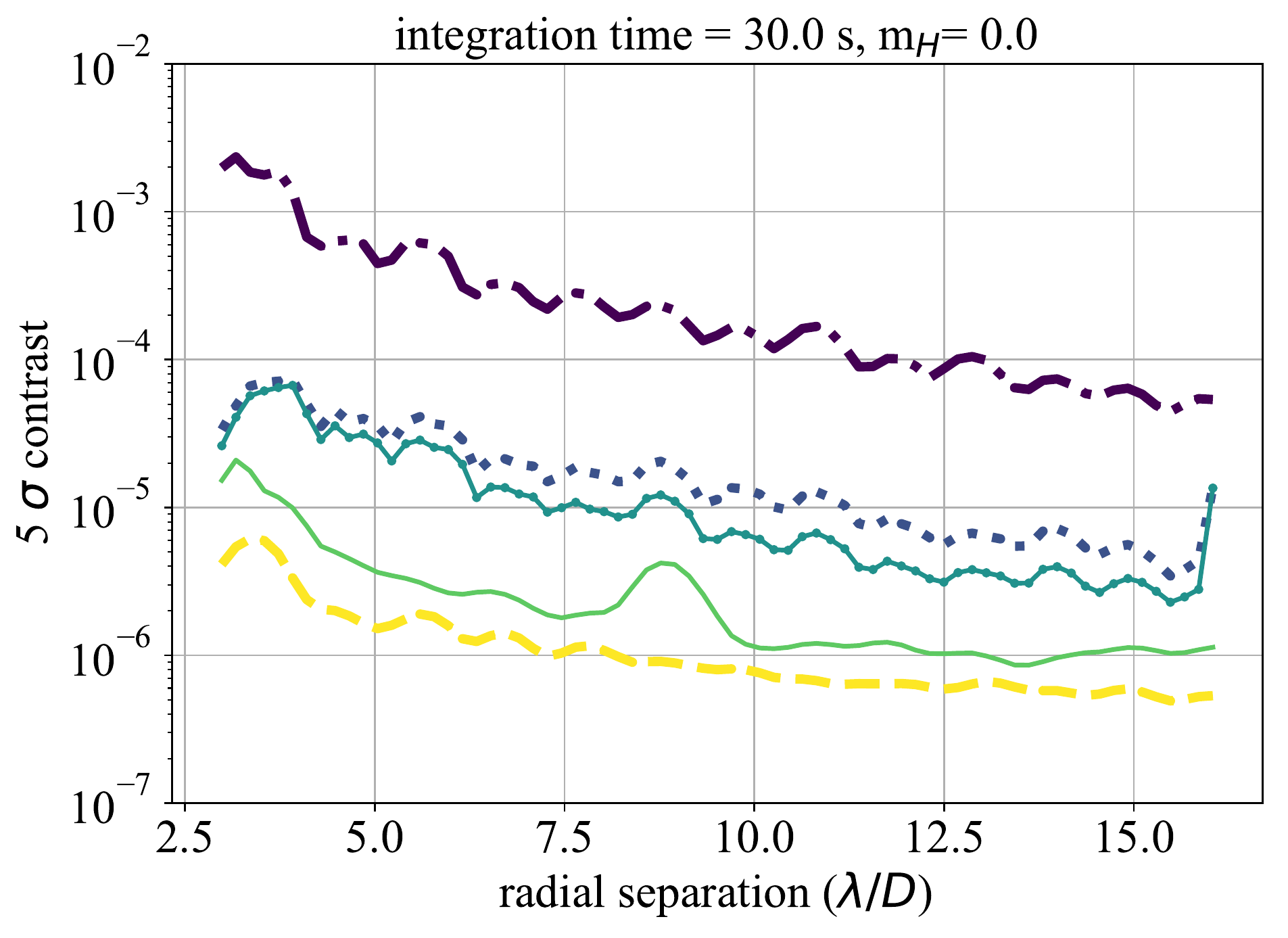}
		\label{fig: b}
	\end{minipage}	
	\begin{minipage}[b]{0.45\textwidth}
		\includegraphics[width=1.0\textwidth]{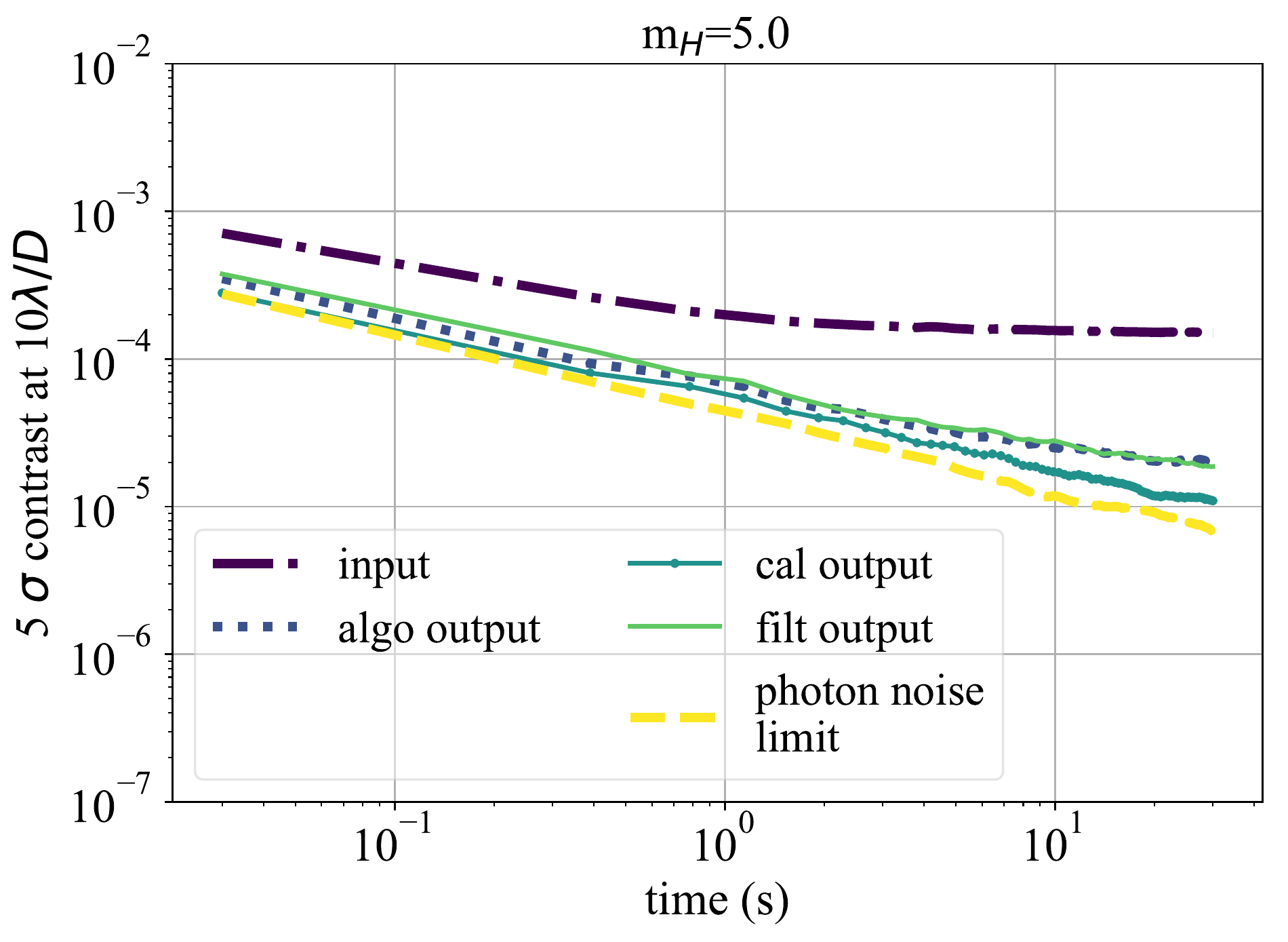}
		\label{fig: d}
	\end{minipage}	
	\begin{minipage}[b]{0.45\textwidth}
		\includegraphics[width=1.0\textwidth]{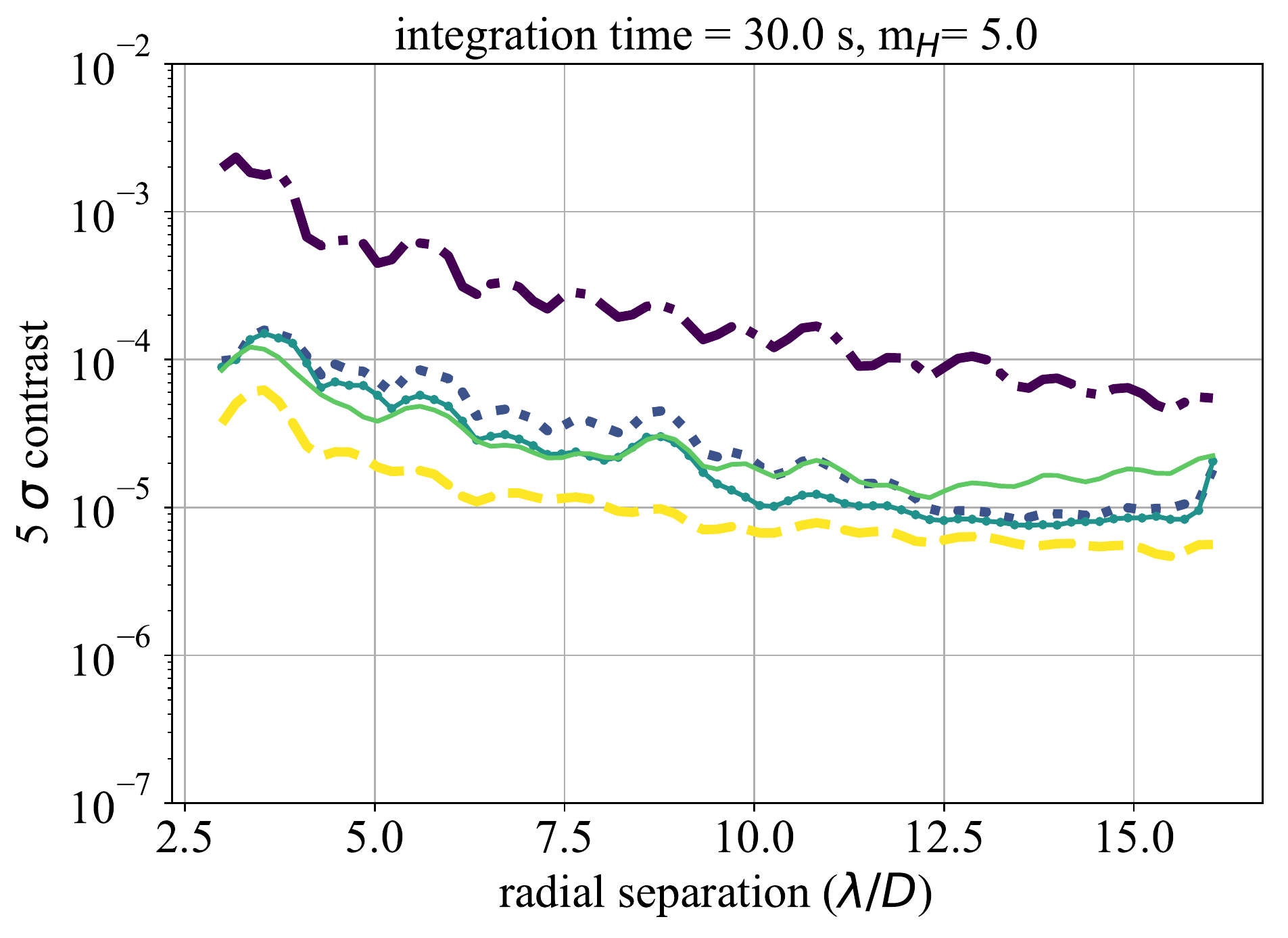}
		\label{fig: d}
	\end{minipage}	
\caption{Simulation results for a 30 second stacked exposure on a 0th and 5th magnitude star (top and bottom rows, respectively). The left and right columns show the contrast at 10 $\lambda/D$ vs. time and full contrast curves after 30 seconds, respectively. The legend in the lower left pannel is the same for all panels. }
\label{fig: stack}
\end{center}
\end{figure}

\begin{figure}[!h]
\begin{center}
\includegraphics[width=1.0\textwidth]{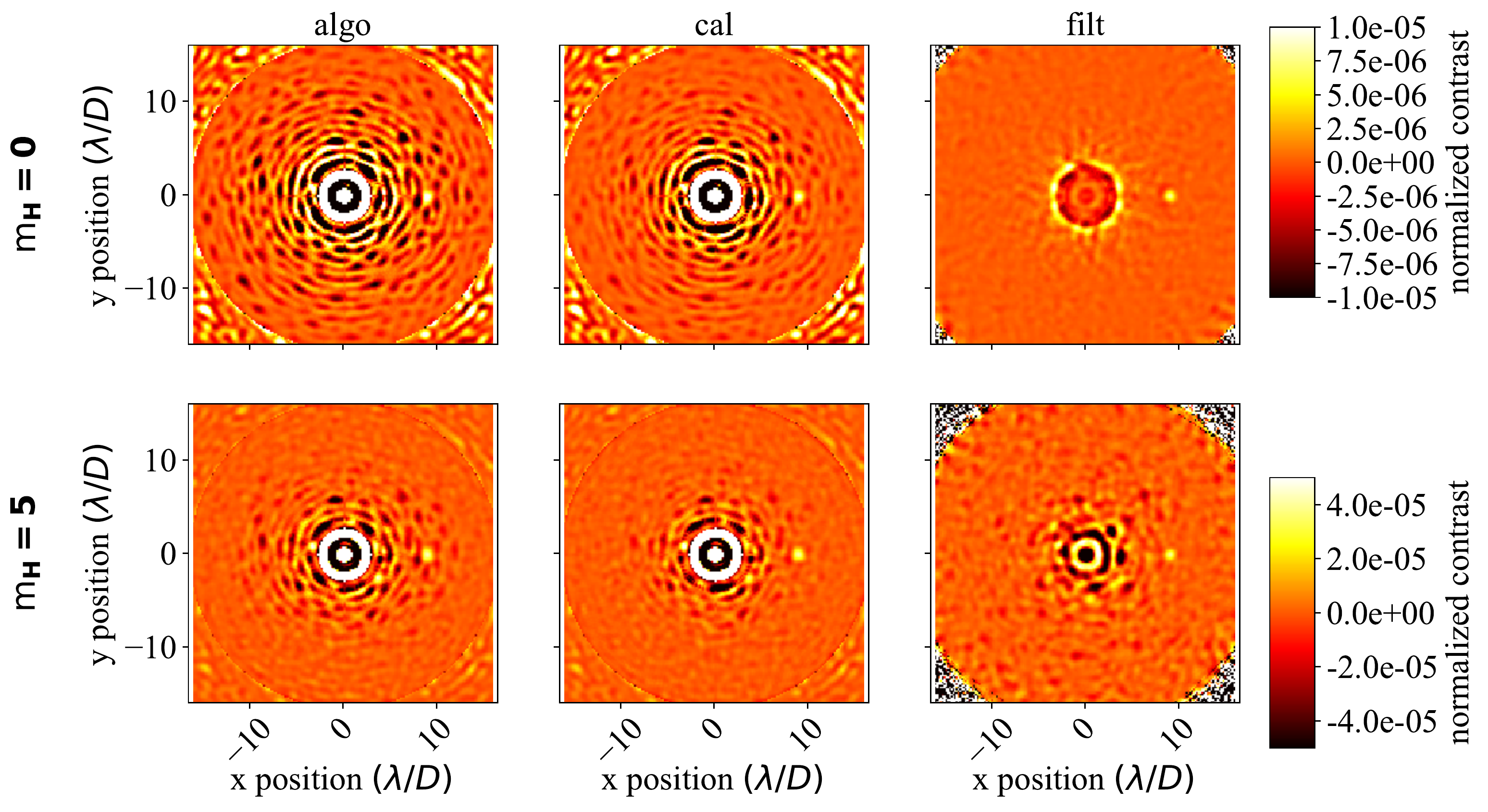}
\caption{The stacked images for 0th and 5th magnitude stars (top and bottom row, repsectively) after 30 seconds from each PSF subtraction method presented in this paper: the also, cal, and filt algorithms (left, middle, and right column, respectively). A exoplanet is simulated in each image at a flux-normalized contrast of $10^{-5}$ and $5\times10^{-5}$ for the 0th and 5th magnitude star, respectively.}
\label{fig: final_imas}
\end{center}
\end{figure}

With the use of this subtraction procedure on individual exposures, Figures \ref{fig: stack} and \ref{fig: final_imas} show the results of averaging out to a cumulative exposure time of 30 seconds. A exoplanet is simulated at $(x,y)=(+9,0)\; \lambda/D$ with a flux-normalized contrast of $10^{-5}$ and $5\times10^{-5}$ for the 0th and 5th magnitude star, respectively. The respective algorithmic exoplanet throughput for the algo, cal, and filt algorithms are 1.18, 1.0, 0.8 for the 0th magnitude star and 0.88, 0.84, 0.77 for the 5th magnitude star.

As predicted in \S\ref{sec: fluxnorm} (Figure \ref{fig: clim}), the contrast curves for the pinhole PSF calibration and reconstruction algorithms flatten out, likely due to static limitations from a combination of how well the pinhole PSF is estimated and how the unrealistic assumptions of azimuthal symmetry in the images impact our flux normalization procedure. The calibration algorithm for the $m_H=5$ simulation does not yet flatten out because the photon noise limit only improves in contrast by a factor of about 10 over 30 seconds, still above the limits illustrated for the $m_H=0$ simulation. The predicted limits from \S\ref{sec: fluxnorm}, Figure \ref{fig: clim} b---the best achievable contrasts for algo and cal of $4\times10^{-6}$ and $2\times10^{-5}$, respectively---are mostly consistent with the results from Figure \ref{fig: stack}---measured 30 s contrasts for algo and cal of $1\times10^{-5}$ and $6\times10^{-6}$, respectively. Interestingly, the addition of atmospheric speckles and/or photon noise decreases the reconstruction algorithm limit by a factor of 2.5 but increases the calibration algorithm limit by a factor of 3.3.

Figure \ref{fig: stack} shows that contrast in the direct pinhole PSF measurement algorithm continues to improve proportional to $t^{-0.5}$ out to 30 seconds in both simulations. This is an important result, and shows that in principle a direct, simultaneous measurement of the full SCC image and pinhole PSF, even if each exposure can only record a few photons, is not limited by static aberration. For example, extrapolating the $m_H=0$ results, 84 hours of telescope time would reach a $5\sigma$ contrast of $10^{-8}$ at 10$\lambda/D$ for a 1\% bandpass.

Running simulations longer than 30 s would not provide additional insight into the physical limitations considered in this paper, since the limiting and continuous effects observed in the upper left panel of Figure \ref{fig: stack} provide sufficient information to predict the behavior out to longer exposure times for either the 0th or 5th magnitude case. These effects are also predicted and discussed in a noiseless analysis of the limitations from static aberration in \S\ref{sec: fluxnorm}. For the 5th magnitude case, the algo and cal algorithms will flatten out at around the same contrast improvement factors as the 0th magnitude case (approximately 12 and 20, respectively; we can already begin to see this flattening for the algo algorithm after 30 seconds in the lower left panel of Figure \ref{fig: stack}) while the filt algorithm will continuously improve within a factor of a few from of the photon noise limit all the way out to the end of the simulation. The same behavior applies to the 0th magnitude case, for which all three subtraction algorithms clearly predict continuously flat (for algo and cal) or improving proportional to $t^{-0.5}$ (for filt) performance out to any exposure time.
\section{Summary \& Conclusions}
\label{sec: conclusion}
In this paper we have presented a new novel approach to coherent differential imaging (CDI) on ground-based telescopes using the self-coherent camera (SCC), called Fast Atmospheric SCC Technique (FAST). Our main findings are as follows:
\begin{enumerate}
\item We design a new FPM specialized for our FAST approach that allows fringes to be detected in exposures that are of order a few milliseconds (\S\ref{sec: scc_fpm}).
\item We illustrate that the standard approach to ground-based CDI (i.e., taking a long exposure to average out atmospheric speckles and measure only the remaining static speckles) is limited in achievable contrast by residual atmospheric speckles that are above the photon noise limit but not measureable/correctable during long exposures (\S\ref{sec: long_exp}).
\item Developing on the framework of \citet{baudoz_psfsubt}, we show that a post-processing algorithm can be used to reconstruct and subtract both atmospheric and static components of an on-sky SCC image without subtracting an incoherent, off-axis exoplanet, but that this process benefits from additional knowledge of the SCC pinhole PSF (\S\ref{sec: est_pin_psf})
\item We propose three different algorithms to estimate the missing pinhole PSF information:
	\begin{enumerate}
	\item	a direct, ``live'' simultaneous measurement of the pinhole PSF along with the SCC image (\S\ref{sec: filt}),
	\item use of a daytime pinhole PSF calibration image (\S\ref{sec: cal}), and
	\item a pinhole PSF reconstruction algorithm that uses only the recorded SCC image and a series of filtering algorithms (\S\ref{sec: algo})
	\end{enumerate}
\item In \S\ref{sec: sim}, with the new SCC FPM design from \S\ref{sec: scc_fpm} and post-processing algorithms from \S\ref{sec: est_pin_psf}, we illustrate that obtaining fast exposures on the order of a few milliseconds to freeze the atmosphere is a solution to the long exposure limitations from \S\ref{sec: long_exp}. With this approach, we find that the direct pinhole PSF measurement algorithm continually improves contrast with time close to the photon noise limit all the way out to the end of our 30 second simulations for both a 0th and 5th magnitude star. This results suggests high contrast imaging instruments using this technique would no longer be limited by static or atmospheric aberration, and that ``raw'' contrast would improve simply by integrating longer.
\end{enumerate}

This paper sets the foundation for fast focal plane wavefront sensing with the SCC. Using the FAST framework presented in this paper, future papers will investigate FAST chromaticity, DM control, instrument-specific simulations, and a laboratory demonstration. Addressing these topics are essential before a FAST strategy can be implemented on-sky; we are currently considering FAST applications to a future upgrade of the Gemini Planet Imager (GPI) and/or Subaru Coronagraphic Extreme Adaptive Optics (SCExAO) instrument(s).
\section*{Acknowledgements}
We gratefully acknowledge research support of the Natural Sciences and Engineering Council (NSERC) of Canada through the Postgraduate Scholarships-Doctoral (PGS-D) program and the Technologies for Exo-Planetary Science Collaborative Research and Training Experience (TEPS CREATE) program. The authors wish to thank B. Macintosh, O. Guyon, J. Lozi, P. Pathak, and A. Sahoo for helpful discussions which improved this manuscript. The authors thank the anonymous referee for his or her comments and suggestions that have significantly improved this manuscript.
\appendix
\section{The Self-Coherent Camera}
\label{sec: scc}
\subsection{Principle}
\label{sec: principle}
\begin{figure}[!h]
\centering
\includegraphics[width=1.0\textwidth]{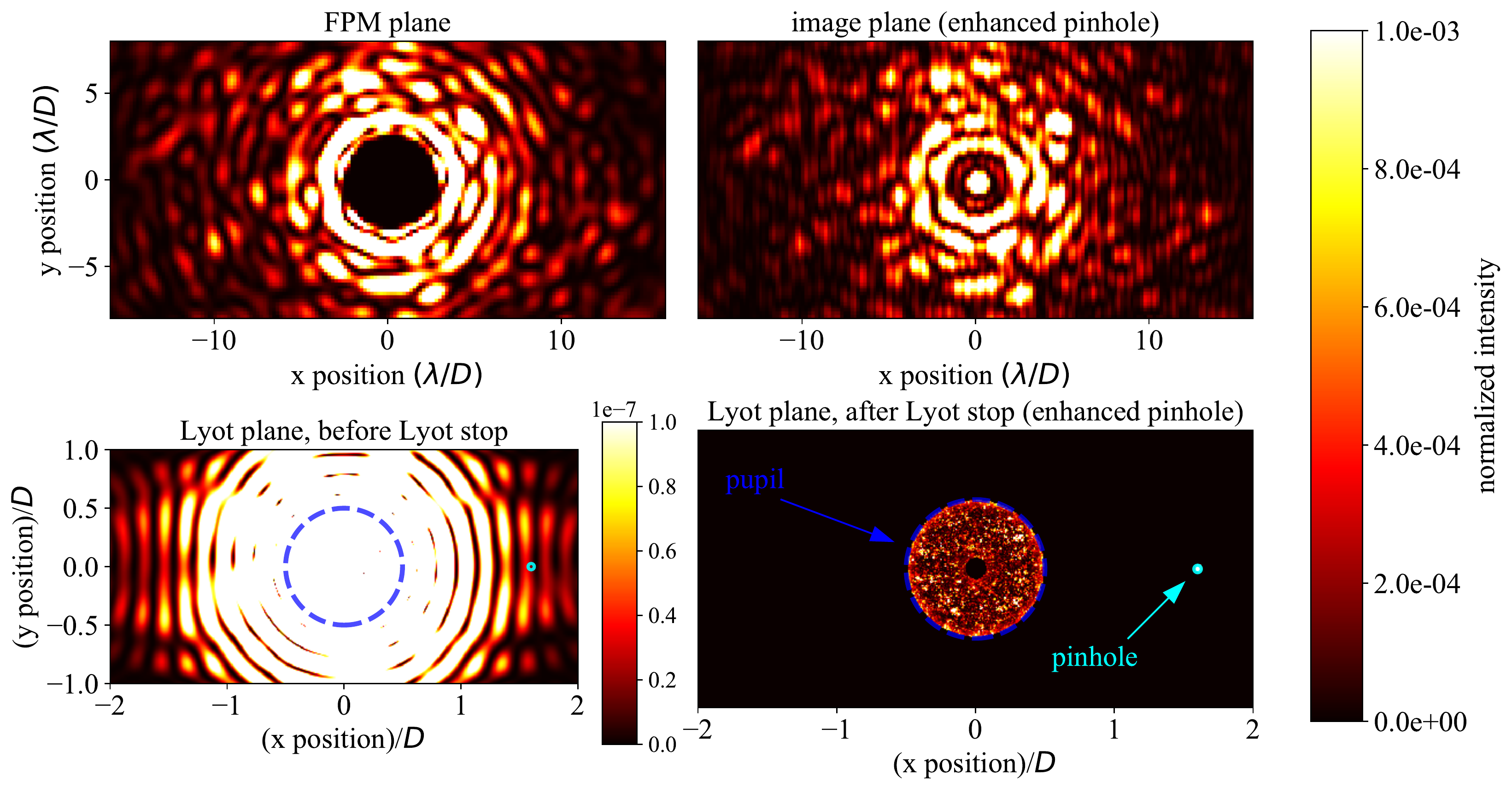}
\caption{Upper left: a FPM is used to occult the PSF core. Lower left: this effect creates diffracted light in the down stream pupil plane at higher spatial frequencies than are present in the original entrance pupil. This effect occurs from any FPM, either in phase and/or amplitude. The contrast scale is different in this image from the other three; the central pupil is saturated to illustrate the relative intensity of this diffraction effect at higher spatial frequencies, of order $10^6$ times dimmer compared to the central pupil. Lower right: a Lyot stop is used to suppress these diffraction effects, but a small off-axis pinhole transmits some of this diffracted light. Upper right: intensity in the downstream image plane after this Lyot stop is applied, showing sub-$\lambda/D$ scale fringes from the interference between the pinhole and main Lyot beam. These fringes provide a direct measurement of the complex electric field in the image plane. The relative intensity in the Lyot pinhole is enhanced by a factor of $2.5\times10^5$ in the upper and lower right images for illustration purposes. All images show the intensity of the complex electric field and are normalized to the peak intensity of the image core if no FPM was present.}
\label{fig: intro}
\end{figure}

The SCC was originally developed by \cite{scc_orig} and then modified by \citet{scc_lyot} specifically to be integrated with high contrast imaging, utilizing the diffraction effects of a coronagraph. The optical effect of adding a hard-edged focal plane mask (FPM), either in amplitude or in phase, to suppress the stellar point spread function (PSF) core is to diffract light to higher spatial frequencies in the down-stream pupil plane than are present in the original entrance pupil. A Lyot stop is then used in this down stream pupil plane in order to suppress such diffraction effects that would ultimately be seen in the coronagraphic image (i.e., the focal plane downstream of the Lyot stop pupil plane). However, the main design difference of the SCC is the additional use of an off-axis pinhole in the Lyot stop. The light transmitted through this pinhole will then interfere with the main Lyot beam in the downstream focal plane, and a camera placed in this focal plane with the necessary sampling will resolve fringes in the image. These recorded fringes retain both phase and amplitude information from the complex electric field of a star. However, no information from the exoplanet is encoded as long as its light does not lie on a discontinuity of the FPM (e.g., on the edge of an amplitude/phase mask). This concept is illustrated in Figure \ref{fig: intro}. With this setup, a recorded monochromatic image at wavelength $\lambda$ is given by \citep{scc_orig}
\begin{equation}
I(\vec{\alpha}) = |A_S(\vec{\alpha})|^2+|A_P(\vec{\alpha})|^2+|A_R(\vec{\alpha})|^2+2\; Re\left\{A_S(\vec{\alpha})A_R(\vec{\alpha})^*\; e^{\left(\frac{2\; i\; \pi \; \vec{\alpha} \; \xi_0}{\lambda}\right)}\right\},
\label{eq: PSF}
\end{equation}
where $A_S(\vec{\alpha}),\; A_P(\vec{\alpha}),\; \text{and } A_R(\vec{\alpha})$ are the focal plane complex electric field components of the star, planet, and light from the Lyot stop pinhole, respectively, $\xi_0$ is the separation between the optical axis and the center of the pinhole in the plane of the Lyot stop, and $\vec{\alpha}$ defines the $(x,y)$ position in the image. The modulation term in equation \ref{eq: PSF}, $(2) Re\left\{A_S(\vec{\alpha})A_R(\vec{\alpha})^*\; e^{\left(\frac{2\; i\; \pi \; \vec{\alpha} \; \xi_0}{\lambda}\right)}\right\}$, is a cosine term with a sub-$\lambda/D$ ($D$ is the diameter of the telescope) spatial scale defined by the separation in the Lyot plane between the main beam and the pinhole. The use of $\vec{\alpha}$ will be removed from subsequent notation in this paper for simplicity, and similarly the modulation term will be replaced by $M (|A_R|\; |A_S|)$, since in this paper we will only be considering speckle subtraction by image processing instead of DM control.

In order to prevent $|A_R|$ from reaching a minimum inside the region of the image correctable by a DM, which would attenuate most of the fringes at that separation and prevent a measurement/correction at sufficient signal-to-noise ratio (SNR), the maximum pinhole diameter, $d$, is \citep{mazoyer}
\begin{align}
d &\leq 1.22 \sqrt{2} \; D/N_\text{act},
\label{eq: pinhole_size}
\end{align}
where $N_\text{act}$ is the number of DM actuators across the pupil. This region of the image correctable by a DM will hereafter be referred to as the ``AO control region.'' In general, equation \ref{eq: pinhole_size} is only a limitation for the area of interest that we want to correct with a DM; a subtraction via image processing may require either a smaller or larger pinhole (to correct for a larger or smaller area in the focal plane, respectively, than the AO control region).
\subsection{Post-Processing Algorithms}
\label{sec: post_processing}
The modulation term from equation \ref{eq: PSF} can be algorithmically isolated in spatial frequency to prevent any loss of exoplanet light. The full processing algorithm to obtain this information from the recorded image is described in \citet[][and references therein]{scc_lyot} and illustrated in Figure \ref{fig: wfsing_algos} (a) and briefly summarized below:
\begin{figure}[!h]
	\begin{minipage}[b]{0.45\textwidth}
		\raggedleft
		\includegraphics[width=1.0\textwidth]{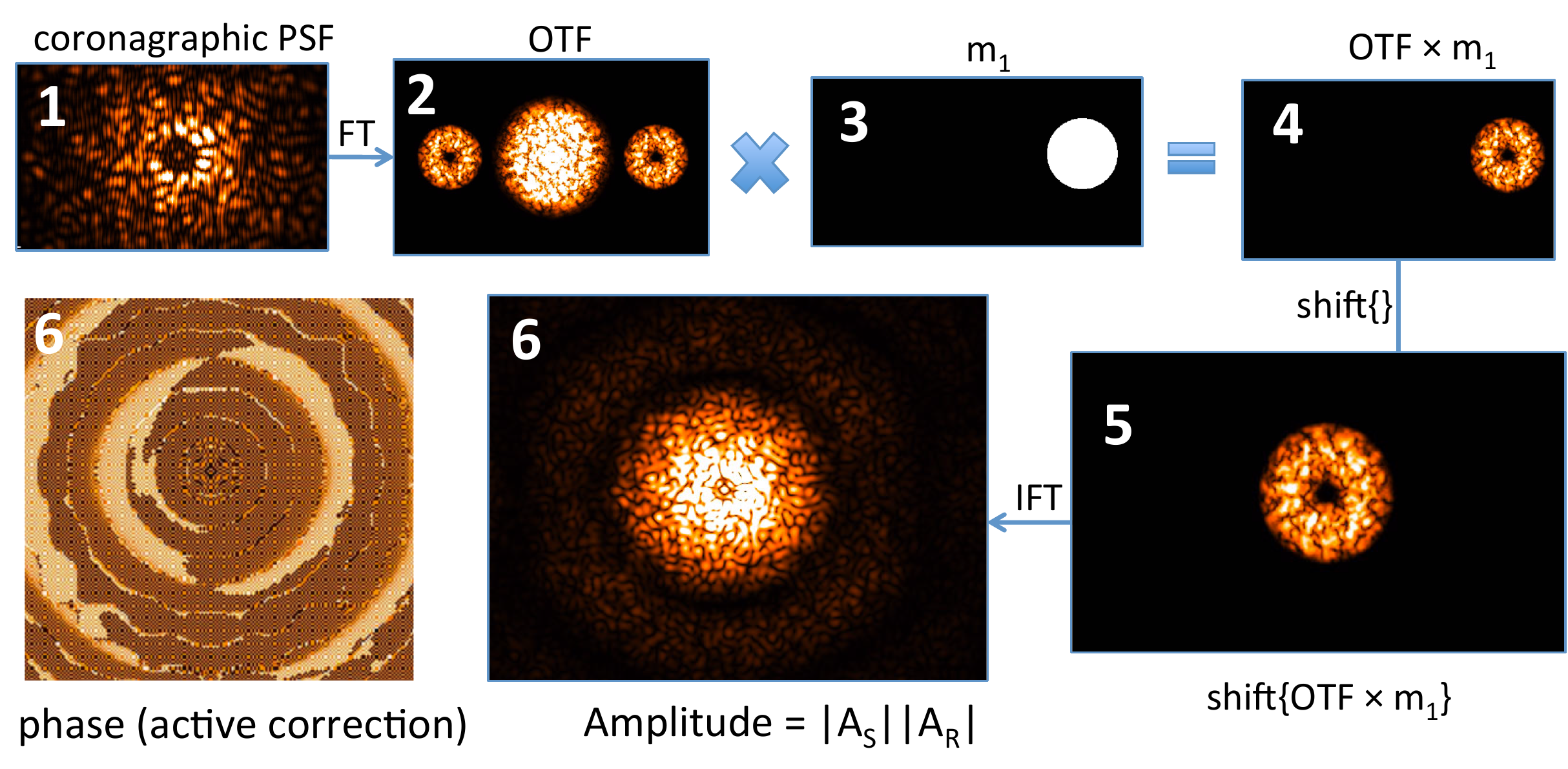}
		\begin{center}
		(a)
		\label{fig: a}
		\end{center}
	\end{minipage}
	\hspace{2cm}\begin{minipage}[b]{0.42\textwidth}
		\raggedright
		\includegraphics[width=0.93\textwidth]{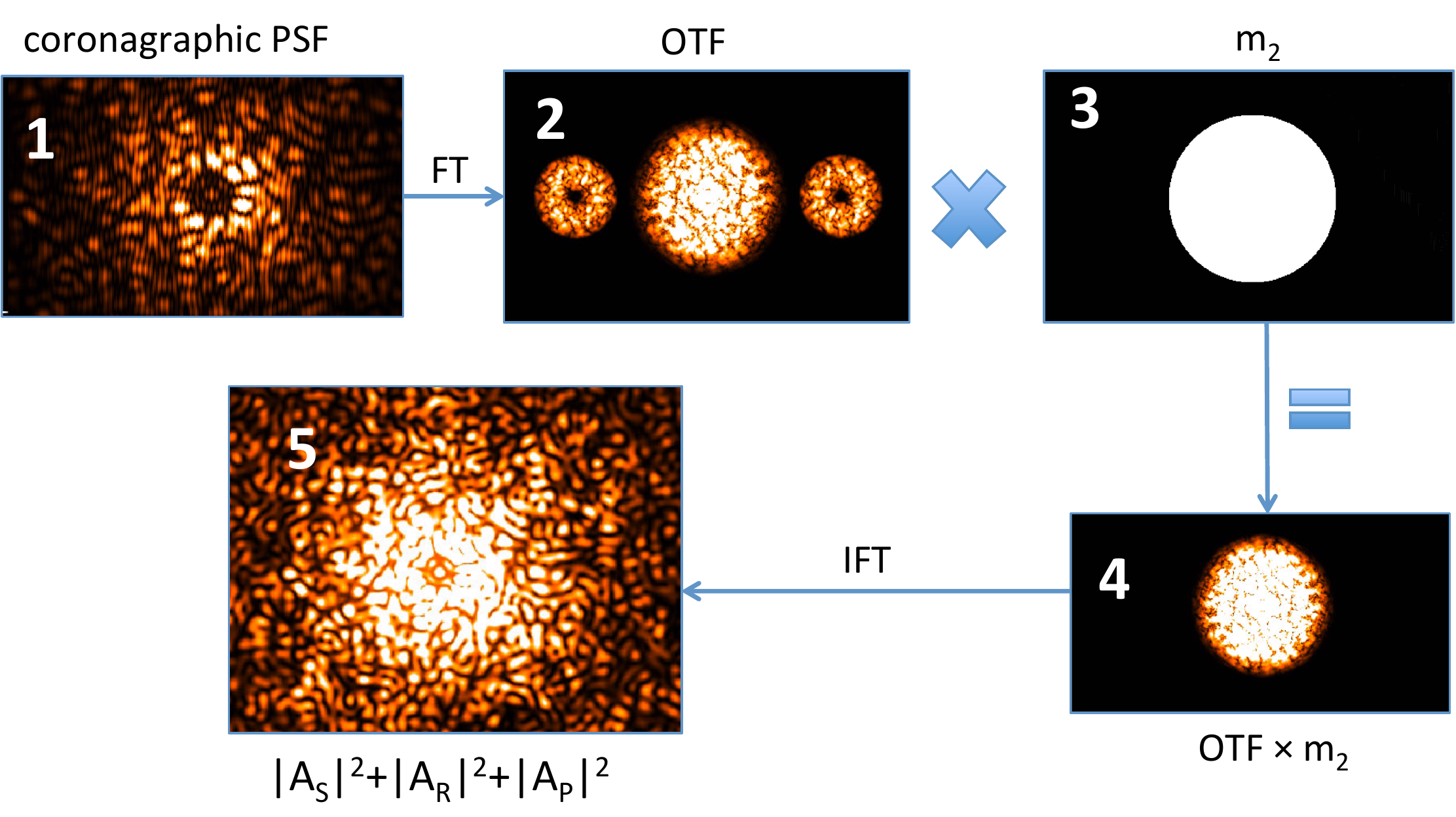}
		\begin{center}
		(b)
		\label{fig: b}
		\end{center}
	\end{minipage}
\caption{Two different Fourier filtering algorithms designed to isolate different SCC image terms in spatial frequency. (a) From \cite{scc_orig}, a Fourier transform of the image (the optical transfer function, or OTF) yields (in amplitude) a main beam and two higher spatial frequency side-lobes, the latter of which represents the electric field of the SCC image modulation amplitude and can be isolated as illustrated here. (b) A separate but similar Fourier filtering algorithm to isolate the un-fringed image components.}
\label{fig: wfsing_algos}
\end{figure} 
\begin{enumerate}
\item taking the Fourier transform of the recorded image to obtain the complex-valued optical transfer function (OTF),
\item multiplying the OTF by a binary mask, $m_1$, to isolate an OTF side lobe, generated from smaller-than-$\lambda/D$ spatial scales in the image,  that contains only the fringed starlight. In order for this side lobe to not overlap in the OTF plane with the main beam, the center of the pinhole in the Lyot plane must be placed at minimum separation, $\xi_0$, from the center of the pupil of \citep{scc_lyot}
\begin{equation}
\xi_0\geq 1.5\;D+d/2
\label{eq: pin_sep}
\end{equation}
\item shifting the unmasked complex array to the center of the image, and then 
\item\label{step: I_minus} applying an inverse Fourier Transform back to the image plane. 
\end{enumerate}
This post-processing algorithm, or Fourier sidelobe isolation algorithm, is also the standard wavefront sensing algorithm used for DM control and is crucial to preserving optical exoplanet throughput. As long as no exoplanet light is sent through the reference pinhole, all of the exoplanet information will reside in the main beam of the OTF, whose amplitude is shown in Figure \ref{fig: wfsing_algos}; by isolating the OTF sidlobe, we are directly measuring the electric field of the star without confusion/contamination from the exoplanet.  

The product produced from step \ref{step: I_minus} above, referred to in the literature as $I_-$, is an algorithmic solution to isolate the complex amplitude of the modulation term in equation \ref{eq: PSF}. By masking only one side lobe in the OTF, this removes both the factor of two in the modulation term and the modulated component of that term, $M$, since all other baselines are also removed (shifting the OTF side-lobe to the center of the Fourier plane after applying a binary mask will only change the phase of $I_-$ in the focal plane, but not the amplitude). Thus, taking the absolute value of $I_-$ gives
\begin{equation}
\Iminus= |F^{-1}\{F\{I\} \; m_1\}|= |A_S|\; |A_R|.
\label{eq: I_minus_a}
\end{equation}
The phase term from this filtering algorithm is used for DM control in a least-squares approach by \cite{mazoyer} and subsequent papers, but it will not be used in this paper for our PSF subtraction approach.

Another image term can be separately reconstructed by placing a different binary mask in the OTF plane, illustrated in Figure \ref{fig: wfsing_algos} (b): removing the modulation term by isolating the main beam with an amplitude mask, $m_2$, then doing an inverse Fourier transform and taking the absolute value provides the un-modulated terms of the image:
\begin{equation}
F^{-1}\{F\{I\} \; m_2\} = |A_S|^2+|A_R|^2+|A_P|^2.
\label{eq: im_a}
\end{equation}
No absolute value or real component operator is needed on the output of equation \ref{eq: im_a}; $F^{-1}\{F\{I\} \; m_2\}$ is entirely a real image, since the Fourier filtering mask $m_2$ is centro-symmetric.
As discussed in \S\ref{sec: est_pin_psf}, If there was a way to estimate the pinhole PSF, this would then allow a reconstruction of $|A_S|$ using equation \ref{eq: I_minus_a} and thus a full subtraction of all terms in equation \ref{eq: im_a} other than $|A_P|^2$. We discuss pinhole PSF estimation methods in \S\ref{sec: est_pin_psf}. Here, assuming we have obtained a noiseless image of the pinhole PSF, from the same wavefront realization as equation \ref{eq: im_a}, the subtracted image, $\text{im}_\text{subt}$ is then
\begin{align}
\label{eq: imsubt}\text{im}_\text{subt}&=F^{-1}\{F\{I\} \; m_2\}-\left(\frac{\Iminus^2}{I_R}+I_R\right) \nonumber \\
&=\left(|A_S|^2+|A_R|^2+|A_P|^2\right)-\left(\frac{\Iminus^2}{I_R}+I_R\right) \nonumber \\
&=|A_P|^2+\left(|A_S|^2-\frac{\Iminus^2}{I_R}\right).
\end{align}
Using the simulation parameters described in appendix \ref{sec: setup}, using a single atmospheric and static phase screen realization we obtain a starting 1$\sigma$ contrast across the full $32\times32\; \lambda/D$ AO control region of $6.3\times10^{-4}$. Using this noiseless target image to calculate $\Iminus$ and then, additionally using the noiseless pinhole PSF, the resulting subtracted image from equation \ref{eq: imsubt} reaches a 1$\sigma$ contrast of $2.0\times10^{-18}$ across the full AO control region (when no companion is simulated); this limit is from unphysical numerical noise in our simulations, which illustrates that we will not be anywhere close to seeing these effects once we consider the effects of photon noise, and that instead when analyzing noisy images we are isolating the physical effects of photon noise propagation through our coronagraphic image reconstruction algorithm. Thus, with no photon noise,
\begin{align}
\left(|A_S|^2-\frac{\Iminus^2}{I_R}\right)&=0,\text{ or} \nonumber \\
\label{eq: As_relation} \frac{|F^{-1}\{F\{I\} \; m_1\}|^2}{I_R}&=|A_S|^2
\end{align}
However, in reality photon noise sets a fundamental limit on the contrast we can achieve in equation \ref{eq: imsubt}. These effects will be discussed next in \S\ref{sec: phlim}.
\subsection{Photon Noise}
\label{sec: phlim}
\begin{figure}[!h]
	\begin{minipage}[b]{0.55\textwidth}
		\begin{center}
		\includegraphics[width=1.0\textwidth]{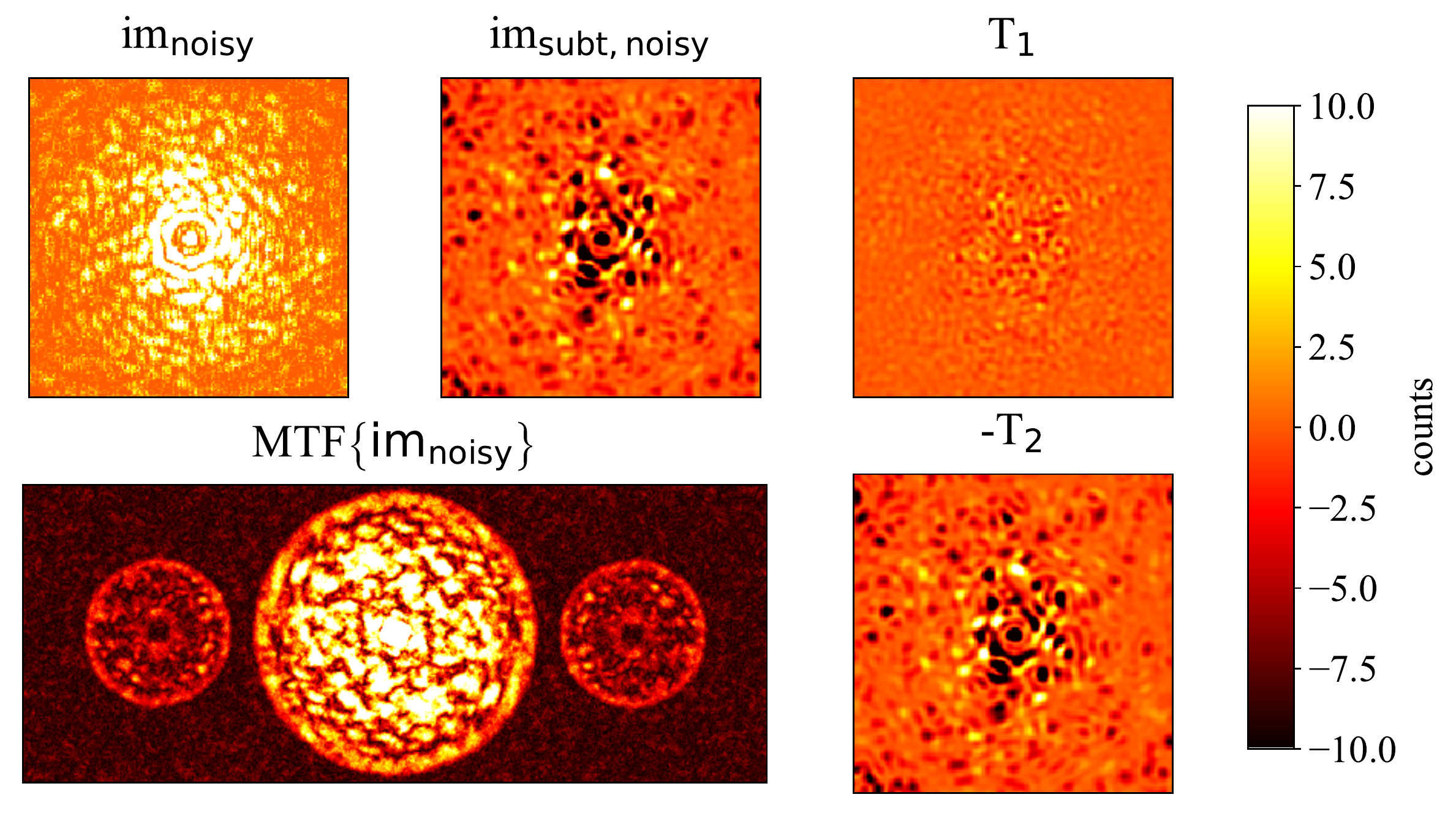}
		(a)
		\label{fig: a}
		\end{center}
	\end{minipage}
	\begin{minipage}[b]{0.4\textwidth}
		\begin{center}
		\includegraphics[width=1.0\textwidth]{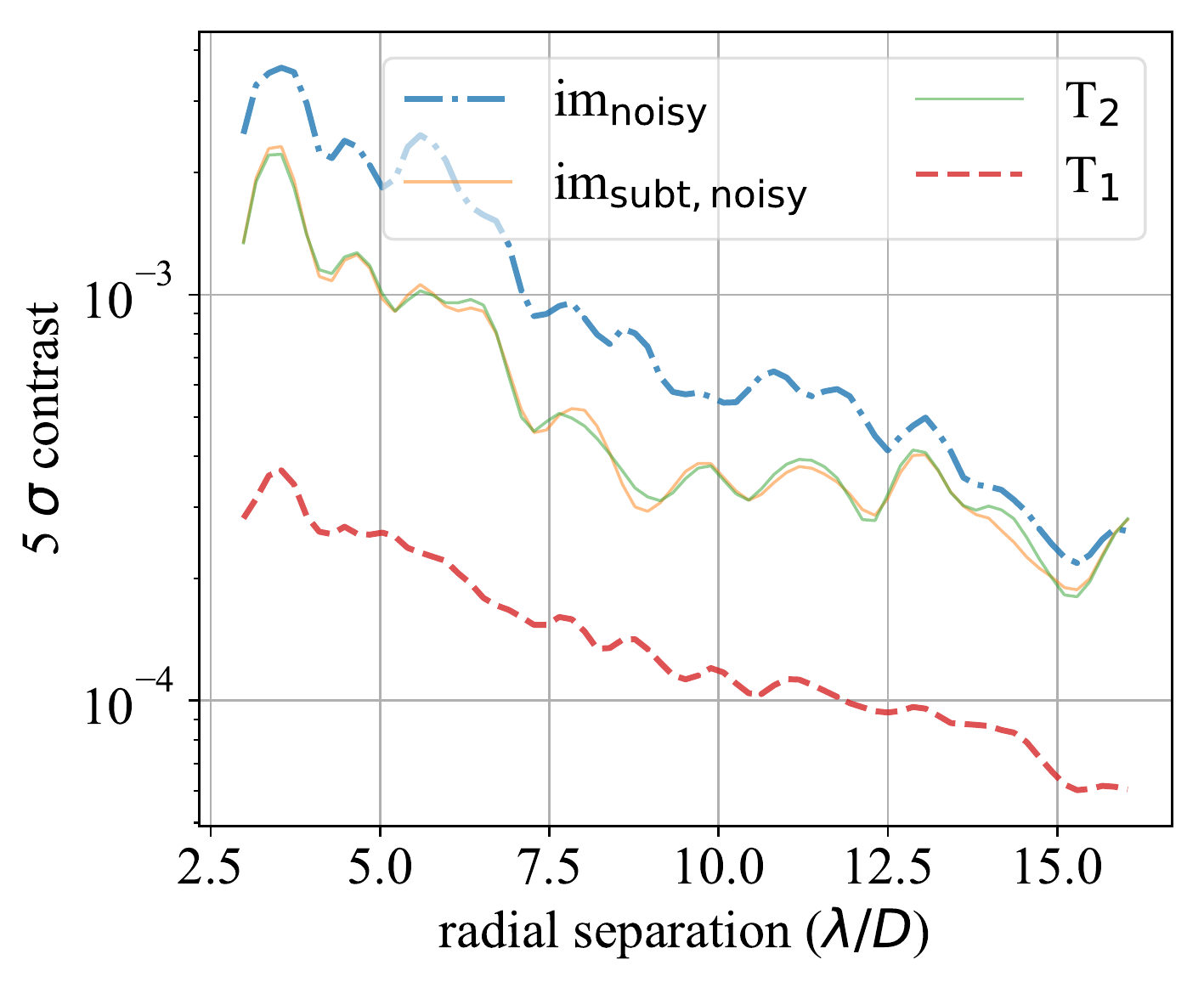}
		(b)
		\label{fig: b}
		\end{center}
	\end{minipage}	
\caption{(a) Assuming perfect knowledge of the pinhole PSF, various terms from equation \ref{eq: noise_propagation} are shown for a 1 ms exposure, with photon noise, on a $m_H=0$ star, with the pinhole intensity enhanced by a factor of $2.5\times10^5$, and with all other parameters are the default values from appendix \ref{sec: setup}: the recorded noisy image (upper left), the subtracted image/photon noise limit (upper middle), the photon noise limit from a noisy SCC image but Fourier filtered to remove fringes so that no spatial scales exist that are smaller than $\lambda/D$ ($T_1$, upper right), and the photon noise limit of the reconstructed $|A_S|^2$ term from a noisy SCC image but Fourier filtered to isolate the modulation amplitude ($-T_2$, lower right). For all these images the FOV is 32$\times$32 $\lambda/D$ and a color bar is shown on the right in illustrating the number of photons recorded and remaining in the subtracted images. The MTF of the upper left image is also shown in the lower left, illustrating the relatively higher impact of photon noise on the sidelobes (i.e., modulation amplitude) compared to the central beam. \\ (b) The contrast curves for the images in Figure (a).}
\label{fig: phlim}
\end{figure}

To understand how photon noise is propagated through the subtraction algorithm in equation \ref{eq: imsubt}, independent of the pinhole PSF estimation, we use the same noiseless $I_R$ but in addition a noisy target image, $\text{im}_\text{noisy}$, to estimate the modulation amplitude, $|I_-|_\text{est}$. By decomposing the noisy image into a noiseless component and a noise-only component (i.e., $\text{im}_\text{noisy}=\text{im}_\text{noiseless}+\left[\text{im}_\text{noisy}-\text{im}_\text{noiseless}\right]$) the subtracted noisy image is then
\tiny
\begin{align}
\label{eq: noise_propagation} \text{im}_\text{subt, noisy}&=F^{-1}\{F\{\text{im}_\text{noisy}\} \; m_2\}-\left(\frac{|F^{-1}\{F\{\text{im}_\text{noisy}\} \; m_1\}|^2}{I_R}+I_R\right) \nonumber \\
&=F^{-1}\{F\{\left(\text{im}_\text{noiseless}+\left[\text{im}_\text{noisy}-\text{im}\right]\right)\} \; m_2\}-\left(\frac{|F^{-1}\{F\{\text{im}_\text{noisy}\} \; m_1\}|^2}{I_R}+I_R\right) \nonumber \\
&=F^{-1}\{F\{\text{im}_\text{noiseless}\} \; m_2\}+F^{-1}\{F\{\text{im}_\text{noisy}-\text{im}_\text{noiseless}\} \; m_2\}-\left(\frac{|F^{-1}\{F\{\text{im}_\text{noisy}\} \; m_1\}|^2}{I_R}+I_R\right) \nonumber \\
&=\left(|A_S|^2+|A_R|^2+|A_P|^2\right)+F^{-1}\{F\{\text{im}_\text{noisy}-\text{im}_\text{noiseless}\} \; m_2\}-\left(\frac{|F^{-1}\{F\{\text{im}_\text{noisy}\} \; m_1\}|^2}{I_R}+|A_R|^2\right) \nonumber \\
&=|A_P|^2+F^{-1}\{F\{\text{im}_\text{noisy}-\text{im}_\text{noiseless}\} \; m_2\}-\left(\frac{|F^{-1}\{F\{\text{im}_\text{noisy}\} \; m_1\}|^2}{I_R}-|A_S|^2\right) \nonumber \\
&=|A_P|^2+F^{-1}\{F\{\text{im}_\text{noisy}-\text{im}_\text{noiseless}\} \; m_2\}-\left(\frac{|F^{-1}\{F\{\text{im}_\text{noisy}\} \; m_1\}|^2-|F^{-1}\{F\{\text{im}_\text{noiseless}\} \; m_1\}|^2}{I_R}\right) \nonumber \\
&\equiv |A_P|^2+T_1-T_2,
\end{align}
\normalsize
where the second to last line of equation \ref{eq: noise_propagation} uses the numerical result from equation \ref{eq: As_relation}. Equation \ref{eq: noise_propagation} illustrates that the subtraction algorithm presented above (when using a simultaneous noiseless pinhole PSF and if no exoplanet is simulated) \textit{is} the photon noise limit and that two different terms contribute to this limit: 
\begin{enumerate}
\item $T_1$ is the photon noise limit from a noisy SCC image but Fourier filtered to remove fringes according to Figure \ref{fig: wfsing_algos} (b) so that no spatial scales exist in the image that are smaller than $\lambda/D$, and
\item $T_2$ is the photon noise limit of the reconstructed $|A_S|^2$ (i.e., starlight) term from a noisy SCC image but Fourier filtered to isolate the modulation amplitude according to Figure \ref{fig: wfsing_algos} (a).
\end{enumerate}
Images and contrast curves for im$_\text{noisy}$, im$_\text{subt, noisy}$, $T_1$, and $T_2$ are shown in Figure \ref{fig: phlim} (a) and (b), respectively, for a 1 ms exposure on a $m_H=0$ star along with the default parameters from appendix \ref{sec: setup}. A normal amplitude mask FPM and Lyot stop does not diffract enough light into the off-axis Lyot pinhole to see any fringes above the photon noise for a ms exposure, and so, as in Figure \ref{fig: intro}, we unphysically enhance the pinhole intensity in Figure \ref{fig: phlim} by a factor of $2.5\times10^5$. A physical solution to this problem will be presented later in \S\ref{sec: scc_fpm}.

Figure \ref{fig: phlim} illustrates that photon noise in exposures on the order of milliseconds, even on bright stars, sets a fundamental limit on the achievable contrast with this coronagraphic image reconstruction technique. Comparing the two terms that contribute to this photon noise limit, the noise from $T_2$ dominates over the noise from $T_1$, illustrating that higher spatial frequencies of the image at the MTF sidelobe positions are more affected by photon noise (i.e., in terms of contrast) than the central beam of the MTF. This greater impact of photon noise at higher spatial frequencies can be explained by the comparing the ``signal-to-noise ratio'' (SNR) of the MTF sidelobe vs. the central beam in Figure \ref{fig: phlim} (a): although the MTF sidelobes are clearly detected above the background photon noise, the central beam is still significantly brighter than the sidelobes and thus detected at a relatively higher ``SNR.''  Accordingly, this higher spatial frequency, pixel-to-pixel noise in the image ultimately propagates through the $\Iminus$ reconstruction algorithm to $\lambda/D$ spatial-scales in $|I_-|_\text{est}$, causing $\lambda/D$-scale residuals in $T_2$ that dominate over $T_1$ instead of just pixel-to-pixel dominated photon noise. Ultimately, the relative impact of $T_1$ vs. $T_2$ will depend directly on the modulation amplitude and thus the intensity through the Lyot stop off-axis pinhole (i.e., if the MTF sidelobes in Figure \ref{fig: phlim} (a) were brighter, or detected at a higher ``SNR,'' $T_2$ would have a relatively lower contribution to the overall $T_1-T_2$ photon noise limit). At the moment this pinhole light is enhanced by a factor of 2.5$\times10^5$ (in intensity) for illustration, but this is indeed set to a similar value as provided by the physical solution presented in \S\ref{sec: scc_fpm}. Thus, the illustration here showing that the noise from $T_2$ dominates over $T_1$ will also hold throughout this paper.

Figure \ref{fig: phlim} (b) and equation \ref{eq: noise_propagation} show that a direct measurement of the pinhole PSF at infinite SNR would allow subtraction to the photon noise limit (by definition), since $\text{im}_\text{subt, noisy}=T_1-T_2$ (if no exoplanet is simulated). With a noiseless pinhole PSF, the photon noise limit is generated entirely from Poisson noise and linear operators, suggesting that the $\text{im}_\text{subt, noisy}$ output of each new recorded image should be completely uncorrelated from one another and that a continuous stacking of these subtracted ms images would improve contrast proportional to $t^{-0.5}$. Thus, $T_1-T_2$ or its filtered version (see appendix \ref{sec: fluxnorm}) is shown in plots as ``the photon noise limit'' throughout this paper. Regarding additional noise propagated through our subtraction algorithms (\S\ref{sec: est_pin_psf}) as the result of an imperfect estimate of the pinhole PSF, looking at the second to last line of equation \ref{eq: noise_propagation}, we can see that as long as our estimate of the pinhole PSF is 
\begin{enumerate}
\item symmetric (in pixel value distribution) around the true noiseless $I_R$ value, and
\item uncorrelated from frame to frame each time a new noisy image is generated,
\end{enumerate}
the contrast should still improve proportional to $t^{-0.5}$ simply by stacking subtracted images, although at a slightly worse contrast from the photon noise limit. However, as in \S\ref{sec: fluxnorm} and \ref{sec: sim}, in some cases one or both of these conditions may not be met, ultimately limiting the achievable contrast even after an infinite exposure time.
\section{Simulation Parameters and Assumptions}
\label{sec: setup}
Throughout this paper, unless explicitly stated, simulations are run at 1.6 $\mu$m, using an 8.2 meter telescope, the Gemini the entrance pupil (with secondary obscuration but no spiders), and a 1 \% bandpass filter; \cite{mazoyer} show that chromaticity effects on the SCC performance should be negligible over this narrow bandwidth and have similar performance to the monochromatic case, and so here we only use the 1 \% bandpass filter for photon counting purposes in an otherwise achromatic simulation. For a Lyot coronagraph we use a 5 $\lambda/D$ diameter amplitude FPM, and a circular Lyot stop, 4\% undersized relative to the entrance pupil; we did not use an apodizer in our simulations. The term ``algorithmic exoplanet throughput'' refers to the ratio of the peak exoplanet flux between the output and input of a PSF subtraction algorithm, with a recorded image as the input and a PSF-subtracted image (i.e., using image processing) as the output; this expression does not refer optical/instrumental throughput of exoplanet light.  We use the Fraunhofer approximation to propagate the electromagnetic field between pupil and focal planes. We assume use of a 32$\times$32 actuator square DM across the pupil to define the AO control region as a $32\times32\; \lambda/D$ box around the optical axis of the image plane. We use a pinhole along the $+x$ axis of the Lyot plane with a relative size (equation \ref{eq: pinhole_size}) of $d/D=1.22 \sqrt{2}/32\approx5.4$\% and relative separation (equation \ref{eq: pin_sep}) of $\xi_0=1.6\; D+d/2$ (we use a factor of 1.6 instead of 1.5 so that the MTF sidelobes are ``comfortably'' not overlapping). We use a beam ratio (image size divided by pupil diameter, or number of pixels per $\lambda/D$) of 5.36. When simulating exposures to estimate photon noise, we calculate flux in number of photons per second for the above parameters from \cite{flux} and then assume a transmission through the atmosphere of 90\%, transmission through the telescope and instrument of 20\%, and detector quantum efficiency of 80\%. In this paper, 5$\sigma$ contrast curves are calculated from an image as follows:
\begin{enumerate}
\item Normalize the image: divide the propagated coronagraphic image by the peak pixel value of the non-coronagraphic image, obtained from the propagated focal plane intensity before applying a FPM but with a Lyot stop placed in the entrance pupil.
\item Calculate the standard deviation of the image output from step 1 in a 0.6 $\lambda/D$ wide annulus centered at radial separation $\rho$, and then multiply the result by 5 to obtain a ``$5\sigma$ contrast.''
\item Repeat step 2 at each radial separation, increasing by 0.2 $\lambda/D$ increments between 3 and 16 $\lambda/D$, and then plot each contrast value vs. each $\rho$ value  to obtain a ``$5\sigma$ contrast curve.''
\end{enumerate}

To simulate quasi-static aberration, we tune the PSD amplitude and power law of both phase and amplitude aberration to match a raw GPI contrast curve of single IFS slice from a stable, 30 second exposure image; in this sequence, consecutive exposures are about 90\% correlated, suggesting that we are seeing mostly static wavefront error and a small amount of residual atmosphere. Additionally, GPI images show no/very little symmetry; as in \citet{marois_phd}, pure phase or pure amplitude aberration should be approximately symmetric in a coronagraphic image for a small amount of aberration (i.e., so that a first order Taylor expansion of the PSF from \citealt{perrin_psf} is a valid approximation), and so the absence of symmetry implies an approximately equal contribution of phase and amplitude aberration to generating quasi-static speckles. With this in mind, we use a 25 nm rms, -1.5 power law phase aberration and a 1\% rms, -2 power law (in intensity) amplitude aberration to generate a static phase and amplitude aberration, and from both of which tip and tilt are subsequently removed by a least-squares subtraction.

To approximate an ExAO system, we use a 100 nm rms phase screen conjugated to the entrance pupil, or by the Mar{\'e}chal approximation a Strehl ratio of 0.87, with a -2 power law, which is typical for AO-corrected residual turbulence (J.P. V{\'e}ran, private communication), subsequently removing tip and tilt to simulate the effect of a low-order wavefront sensor and remove lower order pointing effects. We do not consider the effects of residual uncorrected aberrations (i.e., the phase screen PSD amplitude and shape should be significantly larger, on the order of tens of microns rms, and closer to -11/3, respectively, beyond the spatial frequencies corresponding to the DM control radius) and therefore will not analyze performance beyond the 16 $\lambda/D$ control radius. We use the framework from \citet{sri}, where the $\alpha$ parameter, a dimensionless number between 0 and 1, is used to simulate atmospheric ``boiling,'' or deviating random effects from pure frozen flow; e.g., $\alpha=0.95$ means that 95\% of the current phase screen (or $\sqrt{100^2(0.95)}\approx$97 nm rms) will be translated to the next phase screen realization via a Fourier shift based on the time interval, telescope diameter, and wind speed and direction, whereas as the other $\sqrt{100^2(1-0.95)}\approx$22 nm rms component of the next phase screen realization will be randomly generated, but with the same -2 power law. In this paper we use a windspeed of 10 m/s in the +x pupil direction, 1 ms time intervals, and $\alpha=0.95$ (i.e., 5\% random turbulence is added every ms).
\bibliography{refs}
%
%
%

\end{document}